\documentclass[%
reprint,
superscriptaddress,
 amsmath,amssymb,
 aps,
]{revtex4-1}

\usepackage{graphicx}
\usepackage{float}
\usepackage{braket}
\usepackage{subcaption}
\usepackage{color, ulem}
\usepackage{gensymb}

\newcommand{\VHr}{\ensuremath{V_{\mathrm{H}}(\mathbf{r})}}

\begin{document}

\title{Hartree theory calculations of quasiparticle properties in twisted bilayer graphene}

\author{Zachary A. H. Goodwin}
\affiliation{Departments of Materials and Physics and the Thomas Young Centre for Theory and Simulation of Materials, Imperial College London, South Kensington Campus, London SW7 2AZ, UK\\}
\author{Valerio Vitale}
\affiliation{Departments of Materials and Physics and the Thomas Young Centre for Theory and Simulation of Materials, Imperial College London, South Kensington Campus, London SW7 2AZ, UK\\}
\author{Xia Liang}
\affiliation{Departments of Materials and Physics and the Thomas Young Centre for Theory and Simulation of Materials, Imperial College London, South Kensington Campus, London SW7 2AZ, UK\\}
\author{Arash A. Mostofi}
\affiliation{Departments of Materials and Physics and the Thomas Young Centre for Theory and Simulation of Materials, Imperial College London, South Kensington Campus, London SW7 2AZ, UK\\}
\author{Johannes Lischner}
\affiliation{Departments of Materials and Physics and the Thomas Young Centre for Theory and Simulation of Materials, Imperial College London, South Kensington Campus, London SW7 2AZ, UK\\}

\date{\today}

\begin{abstract}
A detailed understanding of interacting electrons in twisted bilayer graphene (tBLG) near the magic angle is required to gain insights into the physical origin of the observed broken symmetry phases. Here, we present extensive atomistic Hartree theory calculations of the electronic properties of tBLG in the (semi-)metallic phase as function of doping and twist angle. Specifically, we calculate quasiparticle properties, such as the band structure, density of states (DOS) and local density of states (LDOS), which are directly accessible in photoemission and tunnelling spectroscopy experiments. We find that quasiparticle properties change significantly upon doping - an effect which is not captured by tight-binding theory. In particular, we observe that the partially occupied bands flatten significantly which enhances the density of states at the Fermi level. We predict a clear signature of this band flattening in the LDOS in the AB/BA regions of tBLG which can be tested in scanning tunneling experiments. We also study the dependence of quasiparticle properties on the dielectric environment of tBLG and discover that these properties are surprisingly robust as a consequence of the strong internal screening. Finally, we present a simple analytical expression for the Hartree potential which enables the determination of quasiparticle properties without the need for self-consistent calculations. 
\end{abstract}

\maketitle

\section{Introduction}

The discovery of correlated insulator states and superconductivity in magic-angle twisted bilayer graphene (tBLG)~\cite{NAT_I,NAT_S,TSTBLG,SOM,CSCI,IIS,ECNMA,NAT_SS,NAT_MEI,NAT_CO} has generated tremendous excitement and established the rapidly growing field of twistronics~\cite{TT,NAT_WS2WSe2,WSe2_arXiv,tDBLG_CAO,ODFB,NAT_REV,SSCmfb}. Transport experiments have reported energy gaps in the electronic spectrum of tBLG at charge neutrality and when integer numbers of electrons are added to or removed from the moir\'e unit cell~\cite{SOM,IIS,SSCmfb}, while band structure calculations based on the tight-binding or continuum model approach predict the system to be (semi-)metallic~\cite{GBWT,MBTBLG,FBST,LDE,NSCS,PDTBLG,KDP}. This indicates that electron-electron interactions play an important role in tBLG~\cite{EE}.

To understand the properties of interacting electrons in tBLG, many different theoretical approaches have been used. In strong correlation techniques, such as dynamical mean-field theory~\cite{DMFT}, Quantum Monte Carlo~\cite{KVB,IMACP} or exact diagonalization methods~\cite{PCIS}, an effective Hamiltonian for the flat-band electrons is often constructed by adding Hubbard-like interaction terms to a kinetic energy that describes the hopping between flat band Wannier functions~\cite{MLWO,SMLWF,PHD_1,PHD_2}. However, the resulting Hamiltonian is relatively complicated and contains long-ranged hoppings~\cite{MLWO}. Moreover, the accurate determination of the Hubbard parameters is difficult and the construction of flat-band Wannier functions can be hindered by obstructions~\cite{FTBM,SC_WO}. 


In contrast, mean-field treatments of electron-electron interactions are conceptually more straightforward and do not require the construction of flat band Wannier functions. Cea, Walet and Guinea~\cite{H_CM} used Hartree theory within a continuum model of tBLG to calculate band structures and densities of states (DOS) as function of doping and twist angle. They found that the band structure of doped tBLG changes significantly when electron-electron interactions are included, with results that are in qualitative agreement with recent scanning tunnelling spectroscopy (STS) studies~\cite{ECNMA,NAT_SS,NAT_MEI,NAT_CO} which showed that the Fermi level of the doped system can be pinned at the van Hove singularity (VHS). Several groups have also carried out Hartree-Fock calculations of tBLG and studied broken-symmetry phases~\cite{SCHFC,CG_BSHF,CIHF,GSHS,NTSI,URHF}. These calculations are also all based on a continuum theory for the electronic structure of tBLG. While continuum model calculations are numerically very efficient, they typically employ a short-wavelength cutoff for the plane-wave expansion~\cite{H_CM} of the charge density and do not capture the effect of atomic scale Hubbard interactions. 

Klebl and Honerkamp~\cite{LK_CH} carried out atomistic calculations of the spin susceptibility of tBLG with short-ranged atomic Hubbard interactions using the random-phase approximation and found that tBLG inherits magnetic properties from the untwisted bilayer~\cite{ECM,DitBLG}. Rademaker, Abanin and Mellado~\cite{H_AM} used Hartree theory within an atomistic tight-binding model to calculate the charge density, band structure and local density of states (LDOS) in the AA-stacked region of both undoped and hole-doped tBLG at a single twist angle ($\theta=1.05\degree$). They found that electron-electron interactions smoothen the charge density and observed significant changes in the band structure upon doping in qualitative agreement with the continuum Hartree theory calculations of Cea and coworkers~\cite{H_CM}. More recently, Gonz\'{a}lez and Stauber~\cite{TRvC} investigated broken symmetry phases using atomistic Hartree-Fock theory at a single twist angle ($\theta=1.16\degree$), with particular focus on the influence of the dielectric environment on their relative stability.

In this article, we present a systematic study of the effect of long-ranged Coulomb interactions on the band structure, DOS and LDOS as a function of twist angle and doping in tBLG near the magic angle. Specifically, we carry out self-consistent atomistic Hartree calculations. For tBLG suspended in air, we find that electron-electron interactions induce significant changes to the band structure of doped tBLG. In particular, for twist angles near (but not directly at) the magic angle, the partially occupied bands flatten while the unoccupied or fully occupied bands become more dispersive. This explains both the Fermi level pinning and the different shapes of the VHS observed in recent STS experiments. While most STS experiments have focused on the enhancement of LDOS in AA regions of tBLG, we also predict a significant enhancement of the peak in the AB regions, and hypothesize that these peaks are responsible for instabilities to broken-symmetry states even when the system is not at the magic angle. We also study the dependence of the band structure on the dielectric environment and find that the environment-induced changes are relatively small. This is a consequence of the large internal dielectric screening of tBLG. We do not explicitly investigate broken-symmetry states in this work.

\section{Methods}

We study commensurate moir\'e unit cells of tBLG, defined using the convention of Ref.~\cite{LDE}. We start from an AA stacked bilayer and rotate the top layer anticlockwise about an axis that passes through a carbon atom in both layers. The moir\'e lattice vectors are given by $\textbf{R}_{1} = n\textbf{a}_{1} + m \textbf{a}_{2}$ and $\textbf{R}_{2} = -m\textbf{a}_{1} + (n + m) \textbf{a}_{2}$, where $n$ and $m$ are integers and $\textbf{a}_{1} = (\sqrt{3}/2, -1/2)a_{0}$ and $\textbf{a}_{2} = (\sqrt{3}/2, 1/2)a_{0}$ denote the lattice vectors of graphene with $a_{0} = 2.42~\textrm{\AA}$ being the lattice constant of graphene.

At small twist angles, tBLG undergoes significant in-plane and out-of-plane atomic relaxations~\cite{AC,LREBM,CRAC,STBBG,SETLA,LDLE}. We calculate these relaxations using classical force fields: interactions between atoms in the same graphene layer are modelled using the AIREBO-Morse potential~\cite{AIREBO}, whilst the Kolmogorov-Crespi potential~\cite{KC} is used for interactions between atoms in different layers. All relaxations are carried out with the LAMMPS software package~\cite{LAMMPS}. 

To calculate electronic properties of tBLG, we use atomistic Hartree theory and diagonalize the following Hamiltonian 
\begin{equation}
\mathcal{\hat{H}} = \sum_{i}\varepsilon_{i}\hat{c}^{\dagger}_{i}\hat{c}_{i} + \sum_{ij}[t(\textbf{r}_{i} - \textbf{r}_{j})\hat{c}^{\dagger}_{j}\hat{c}_{i} + \text{H.c.}],
\label{eq:H}
\end{equation}
where $\varepsilon_{i}$ and $\hat{c}^{\dagger}_{i}$ ($\hat{c}_{i}$) denote the on-site energy of a carbon atom and the electron creation (annihilation) operator associated with the p$_{z}$-orbital on atom $i$, respectively. The hopping parameters $t(\textbf{r}_{i} - \textbf{r}_{j})$ between atoms $i$ and $j$ are obtained using the standard Slater-Koster rules~\cite{SK}
\begin{equation}
t(\textbf{r}) = V_{pp\sigma}(\textbf{r})\bigg(\dfrac{\textbf{r}\cdot\textbf{e}_{z}}{|\textbf{r}|}\bigg)^{2} + V_{pp\pi}(\textbf{r})\bigg(1 - \dfrac{\textbf{r}\cdot\textbf{e}_{z}}{|\textbf{r}|}\bigg)^{2},
\end{equation}
with $V_{pp\sigma}(\textbf{r}) = V_{pp\sigma}^{0}\exp\{q_{\sigma}(1 - |\textbf{r}|/d_{AB})\}\Theta(R_c-|\mathbf{r}|)$ and $V_{pp\pi}(\textbf{r}) = V_{pp\pi}^{0}\exp\{q_{\pi}(1 - |\textbf{r}|/a)\}\Theta(R_c-|\mathbf{r}|)$ with $V_{pp\sigma}^{0} = 0.48$ eV and $V_{pp\pi}^{0} = -2.7$ eV~\cite{FC,SK,EPG}. Note that $a = 1.397~\textrm{\AA}$ is the carbon-carbon bond length and $q_{\sigma} = 7.43$ and $q_{\pi} = 3.14$~\cite{LDE,NSCS}. Hoppings between carbon atoms whose distance is larger than the cutoff $R_c=20~\textrm{\AA}$ are neglected~\cite{EDS}.

The on-site energy is determined by the Hartree potential energy \VHr\ according to 
\begin{equation}
\varepsilon_i = \int d\textbf{r} \phi_z^2(\textbf{r}-\textbf{t}_i)\VHr,
\end{equation}
where $\phi_z$ denotes the carbon p$_z$-orbital at position $\mathbf{t}_i$ in the first unit cell [note that \VHr\ is periodic in the moir\'e unit cell]. The Hartree potential is obtained from the electron density $n(\mathbf{r})$ and the screened electron-electron interaction $W(\mathbf{r})$ via
\begin{equation}
\VHr = \int d\textbf{r}' W(\textbf{r}-\textbf{r}') [n(\textbf{r}') - n_0(\textbf{r}')],
\label{eq:VH}
\end{equation} 
where $n_0(\mathbf{r})$ is a reference electron density that ensures overall charge neutrality. 

We consider two cases for the screened interaction. One is for tBLG encapsulated by a dielectric substrate with background dielectric constant $\epsilon_{\mathrm{bg}}$ and, hence, a screened interaction given by $W(\mathbf{r})=e^2/(4\pi\epsilon_0 \epsilon_{\mathrm{bg}}|\mathbf{r}|)$. The other is for the case when there is the additional presence of metallic gates on both sides of the dielectric substrate. Assuming that the tBLG lies in the $x$-$y$ plane, the screened interaction in this case is given by
\begin{equation}
W(\textbf{r}) = \dfrac{e^2}{4\pi\epsilon_0\epsilon_\mathrm{bg}}\sum_{m=-\infty}^{\infty} \dfrac{(-1)^{m}}{\sqrt{|\textbf{r}|^2 + (2m\xi)^2}},
\label{WMG}
\end{equation} 
where $\xi$ is the thickness of the dielectric substrate separating tBLG from the metallic gate on each side~\cite{MGS,PHD_3}.

The charge density can be expressed in terms of the Bloch eigenstates $\psi_{n\mathbf{k}}(\mathbf{r})$ (with subscripts $n$ and $\mathbf{k}$ denoting a band index and the crystal momentum, respectively) of the Hamiltonian in Eq.~\eqref{eq:H} according to
\begin{equation}
\begin{split}
n(\textbf{r})& = \sum_{n\textbf{k}} f_{n\textbf{k}} |\psi_{n\textbf{k}}(\textbf{r})|^2 \\  &=\sum_{j}n_j\chi_j(\textbf{r}),
\end{split}
\end{equation}
where $f_{n\textbf{k}}=2\Theta(\varepsilon_\mathrm{F}-\varepsilon_{n\mathbf{k}})$ is the occupancy of state $\psi_{n\mathbf{k}}$ with eigenvalue $\varepsilon_{n\mathbf{k}}$ (where $\varepsilon_\mathrm{F}$ is the Fermi energy), $\chi_j(\textbf{r}) = \sum_\mathbf{R} \phi_{z}^2(\textbf{r}-\textbf{t}_j-\textbf{R})$ (with $\mathbf{R}$ denoting the moir\'e lattice vectors) and $n_j$ is the total number of electrons in the $j$-th orbital. Note that we neglect contributions to the density that result from the overlap of p$_z$-orbitals on different atoms. 

To construct the reference electron density $n_0(\textbf{r})$, we note that the hopping parameters of Eq.~\eqref{eq:H} were obtained from fits to band structures of graphene and untwisted graphene bilayers calculated using density-functional theory (DFT) and, therefore, include the Hartree potential energy of \textit{the uniform system} (when the occupancy of all carbon atoms is equal)~\cite{NSCS,LDE,FC}. To exclude this contribution to $V_H$ in our tBLG calculations, we use the reference density
\begin{equation}
n_0(\textbf{r}) = \bar{n} \sum_j \chi_j(\textbf{r}),
\label{eq:rho0}
\end{equation}
where $\bar{n}$ is the average of $n_j$. The average filling can be expressed as $\bar{n}=1+\nu/N$, where $\nu$ denotes the number of electrons that have been added to or removed from the moir\'e unit cell and $N$ is the total number of states (also atoms in the moir\'e unit cell). 

To obtain a self-consistent solution of the atomistic Hartree equations, we proceed as follows. We first set $\varepsilon_i=0$ and diagonalize the Hamiltonian, Eq.~\eqref{eq:H}, using an $8\times 8$ k-point grid to sample the first Brillouin zone. From the eigenstates without a potential we first calculate $n_j$. Next, we calculate the on-site energies via
\begin{equation}
    \varepsilon_i = \sum_j (n_j - \bar{n}) \sum_\mathbf{R} W_{\mathbf{R}ij},
\end{equation}
where $W_{\mathbf{R}ij}=W(\mathbf{R}+\mathbf{t}_j-\mathbf{t}_i)$. If $\mathbf{R}=0$ and $i=j$, we set $W_{0,ii}=U/\epsilon_\mathrm{bg}$ with $U=17$~eV~\cite{SECI}. This is equivalent to treating $\phi^2_z$ as a delta-function when considering interactions between different atoms. We carry out calculations for both $\epsilon_\mathrm{bg}=1$ (tBLG suspended in air) and $\epsilon_\mathrm{bg}=3.9$ (tBLG sandwiched between hexagonal boron nitride). To converge the sum over moir\'e lattice vectors, we use a $21\times 21$ supercell. 


In each subsequent iteration of the self-consistent cycle, we mix a fraction of the new Hartree potential with the Hartree potential from the previous iteration. A mixing fraction of 0.1, i.e., the addition of 10 percent of the new potential to 90 percent of the potential from the previous iteration, was found to give satisfactory results in most cases. In a few cases, however, smaller values for the mixing fraction were used to improve convergence. Typically, the Hartree potential converges within 60 iterations to an accuracy of better than 0.1~meV for all doping levels and twist angles considered.

In order to calculate the density of states (DOS) per moir\'e cell, we sample the first Brillouin zone using approximately 6,000 k-points and represent the contribution from each energy level as a gaussian. A similar procedure is used for the local density of states (LDOS). Note that we average the LDOS over atoms within a radius of $15~\textrm{\AA}$ [we found that the results do not depend qualitatively on the radius chosen, provided it is larger than the length scale of the carbon-carbon bond length $\mathcal{O}(1~\textrm{\AA})$ and smaller than the moir\'e length scale $\mathcal{O}(10~\textrm{nm})$]. 

\begin{figure*}[t!]
\centering
\begin{subfigure}{0.325\textwidth}
  \includegraphics[width=1\linewidth]{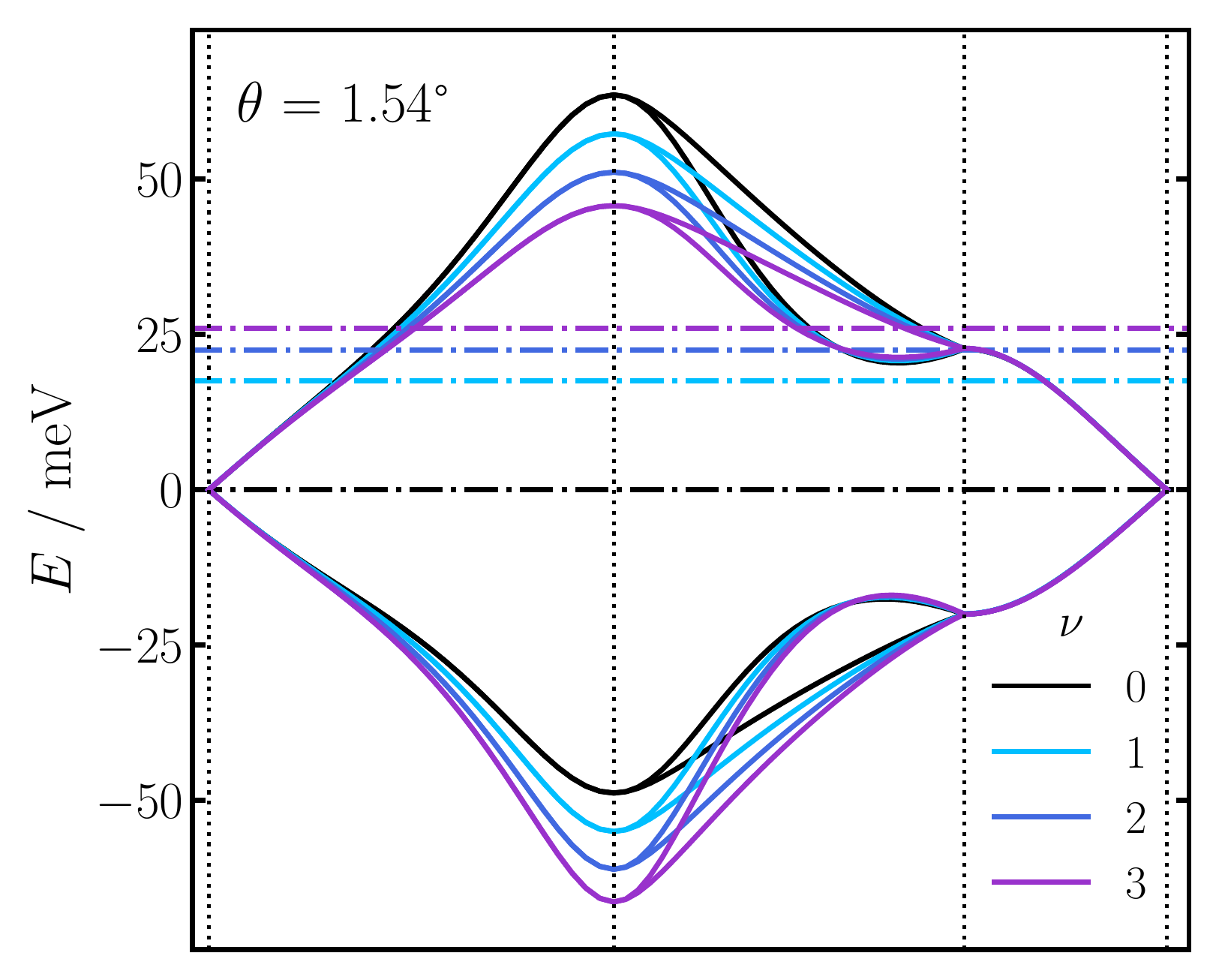}
\end{subfigure}
\begin{subfigure}{0.3\textwidth}
  \includegraphics[width=1\linewidth]{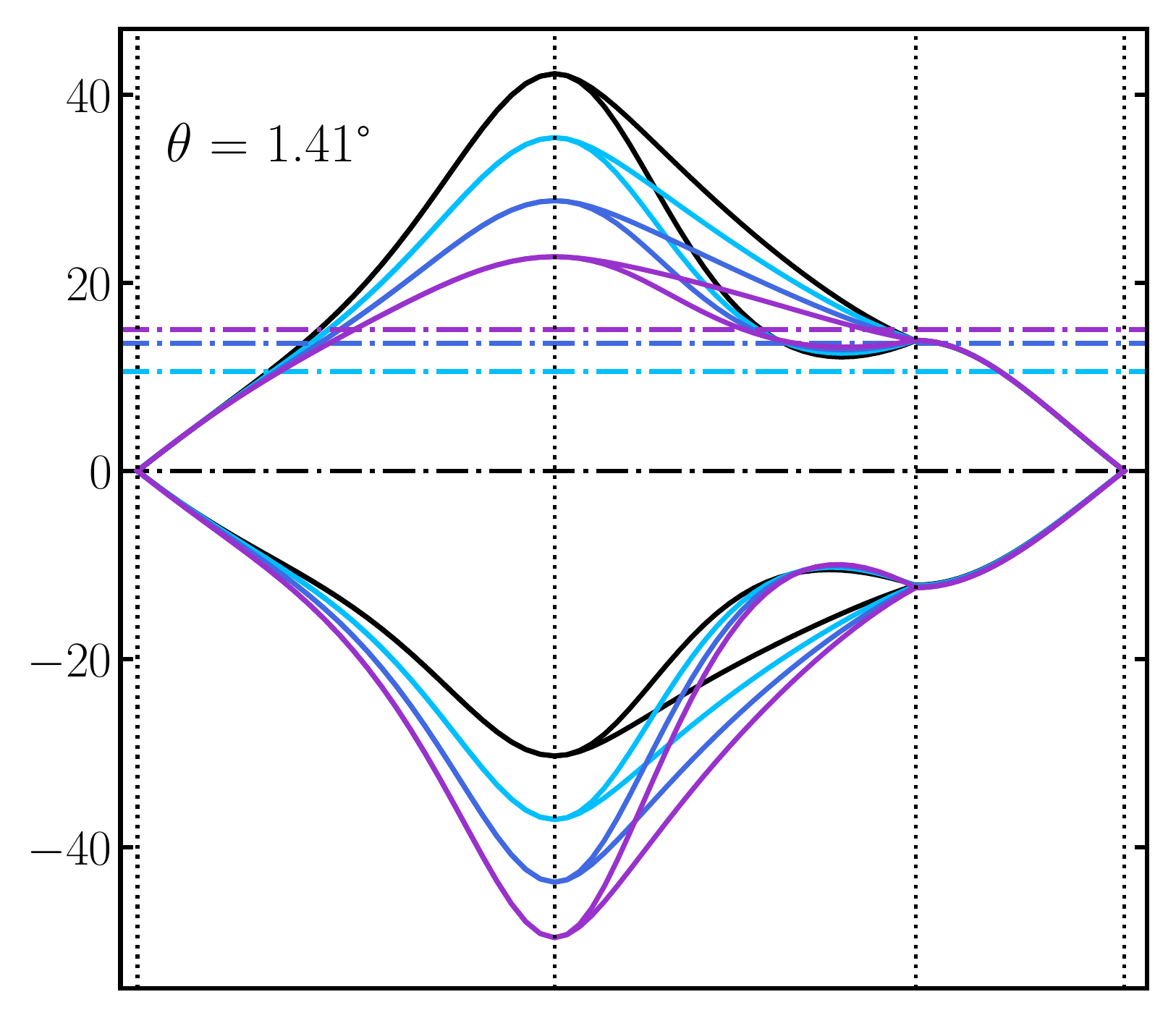}
\end{subfigure}
\begin{subfigure}{0.3\textwidth}
  \includegraphics[width=1\linewidth]{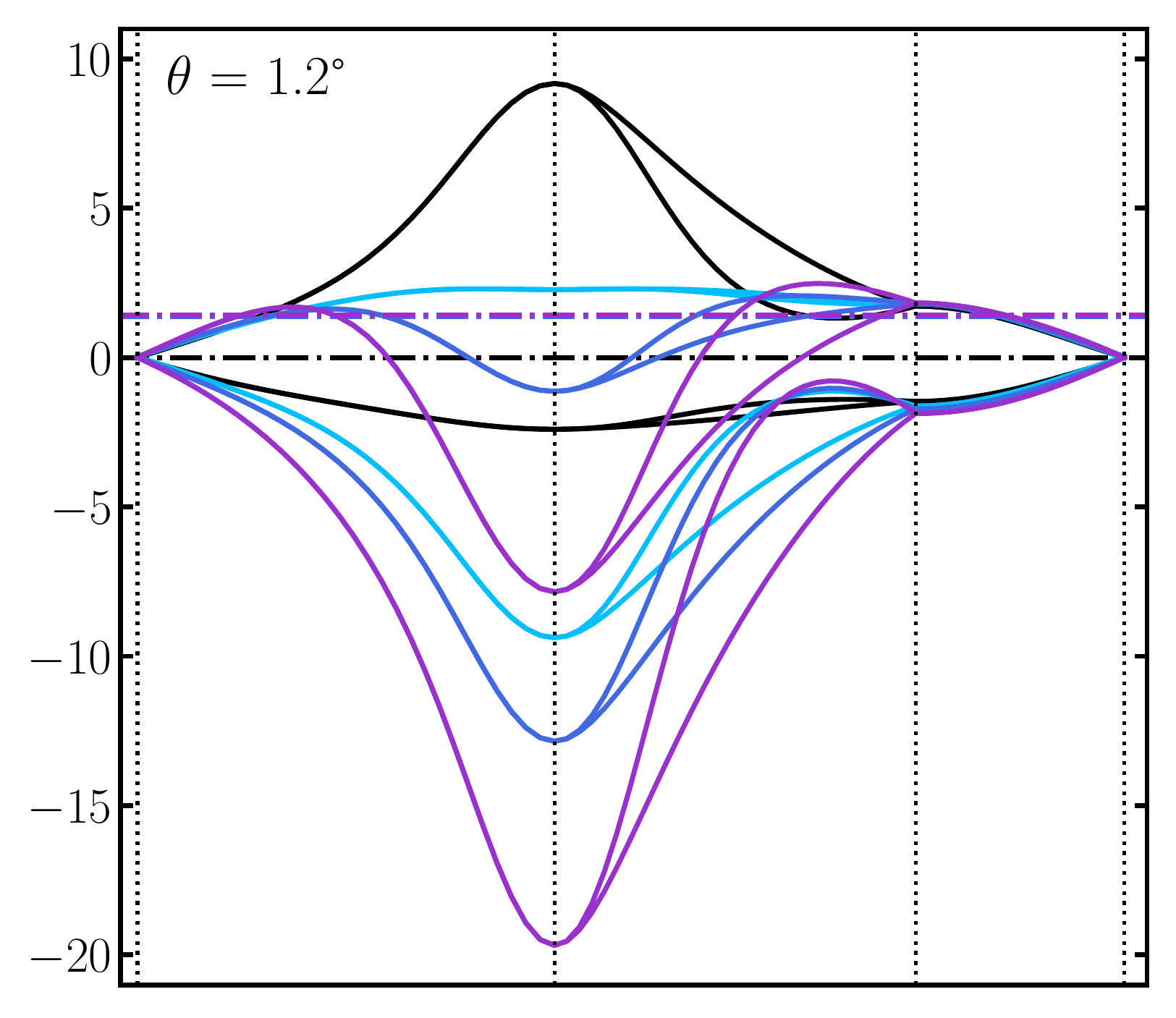}
\end{subfigure}
\begin{subfigure}{0.325\textwidth}
  \includegraphics[width=1\linewidth]{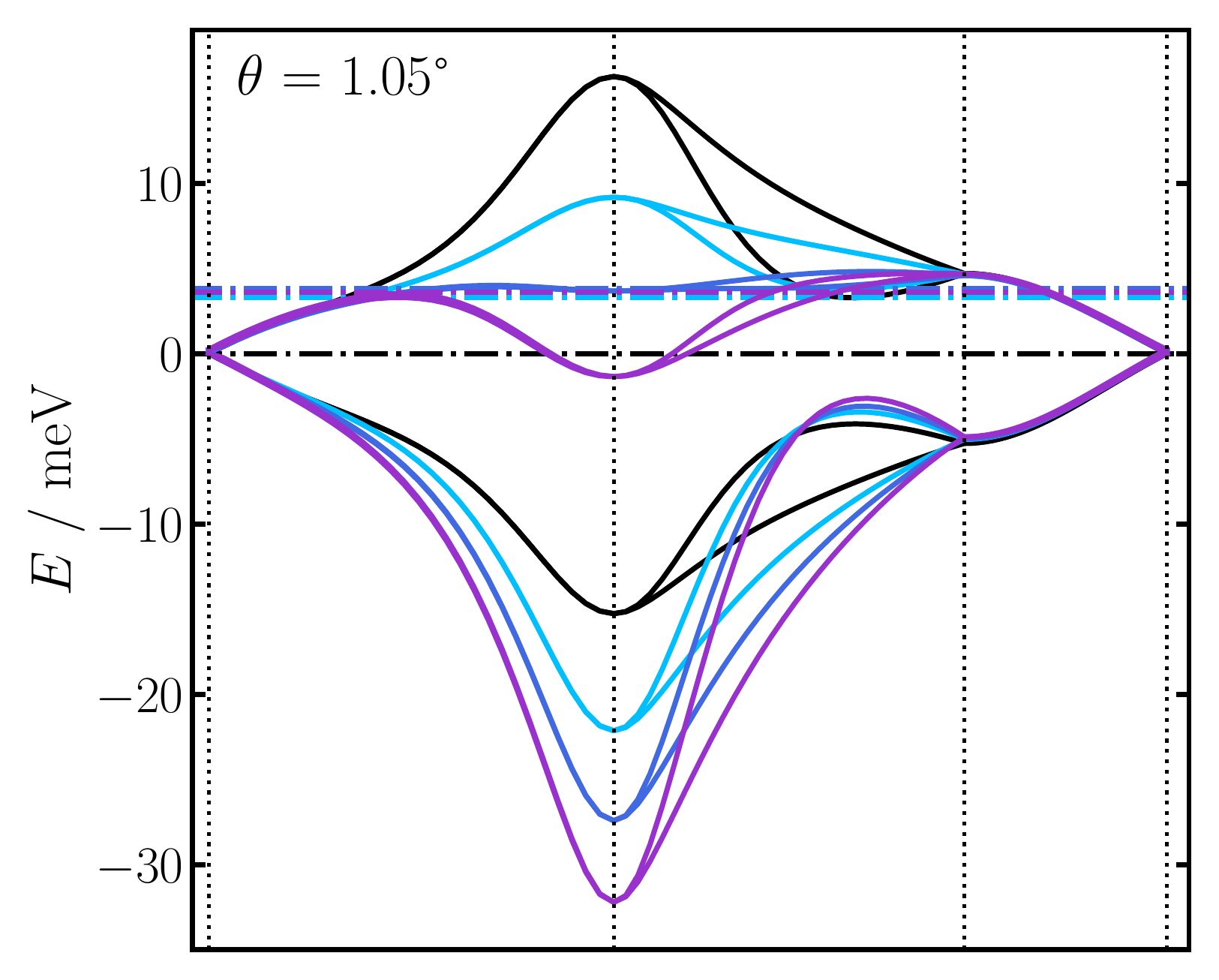}
\end{subfigure}
\begin{subfigure}{0.3\textwidth}
  \includegraphics[width=1\linewidth]{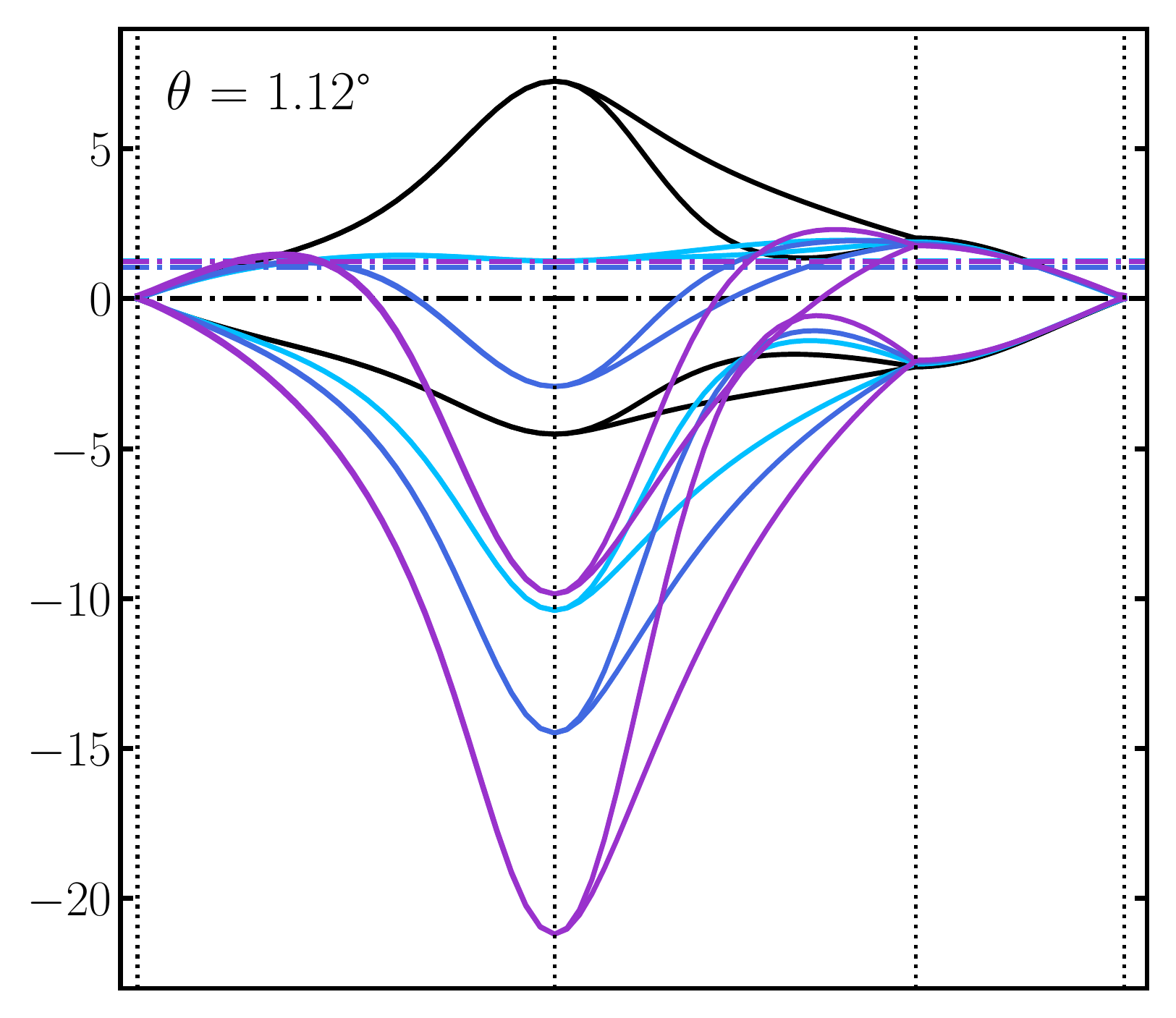}
\end{subfigure}
\begin{subfigure}{0.3\textwidth}
  \includegraphics[width=1\linewidth]{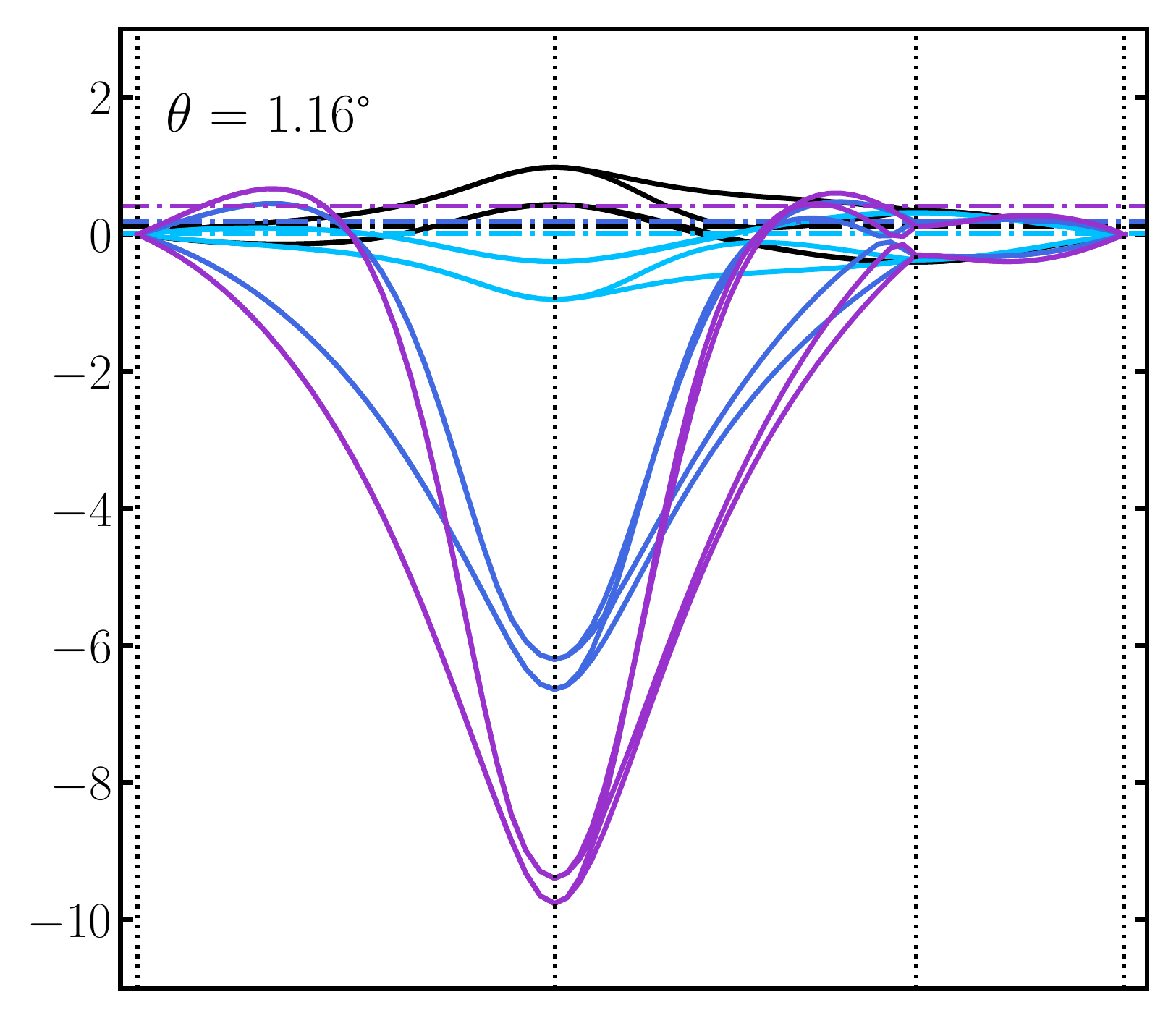}
\end{subfigure}
\begin{subfigure}{0.325\textwidth}
  \includegraphics[width=1\linewidth]{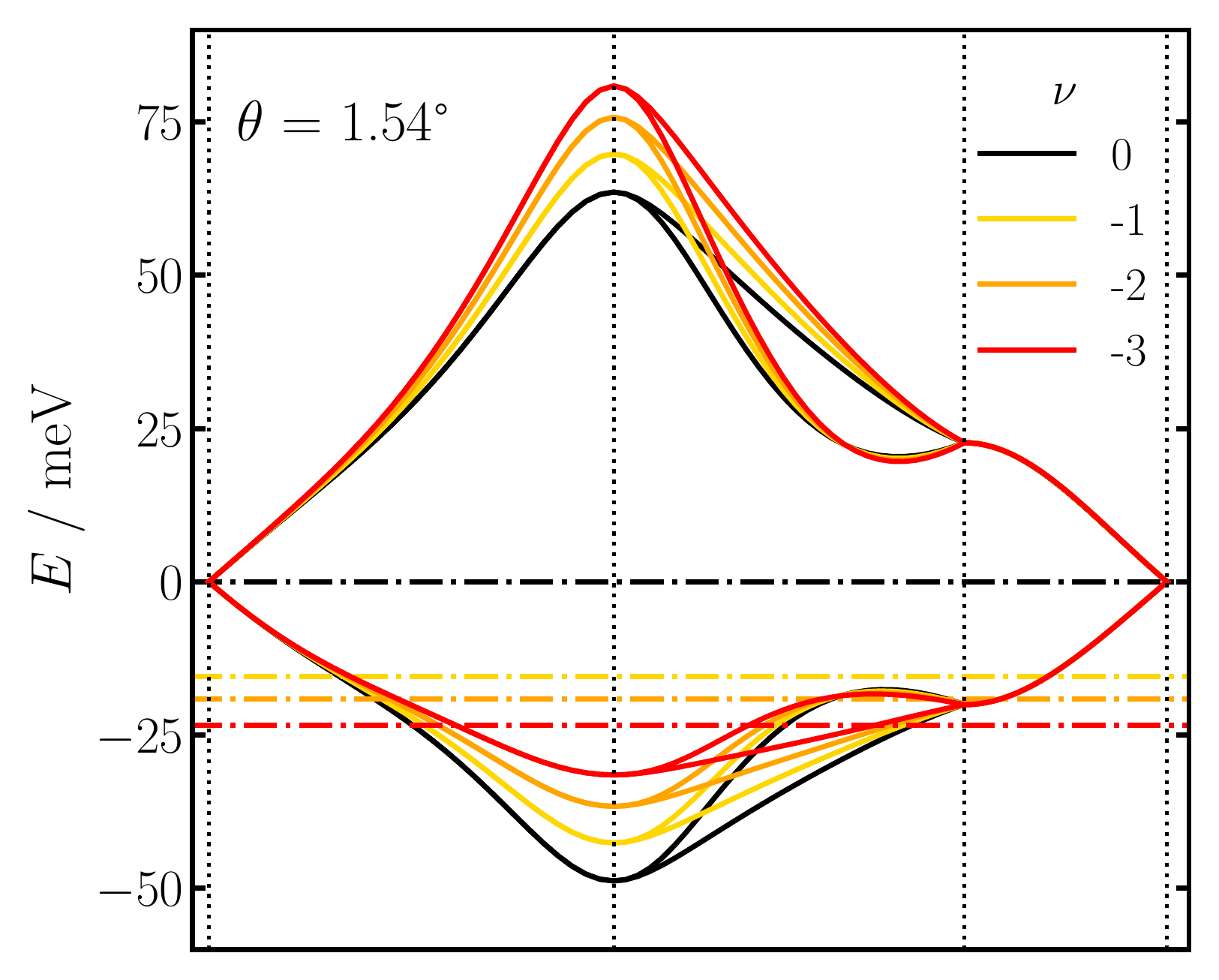}
\end{subfigure}
\begin{subfigure}{0.3\textwidth}
  \includegraphics[width=1\linewidth]{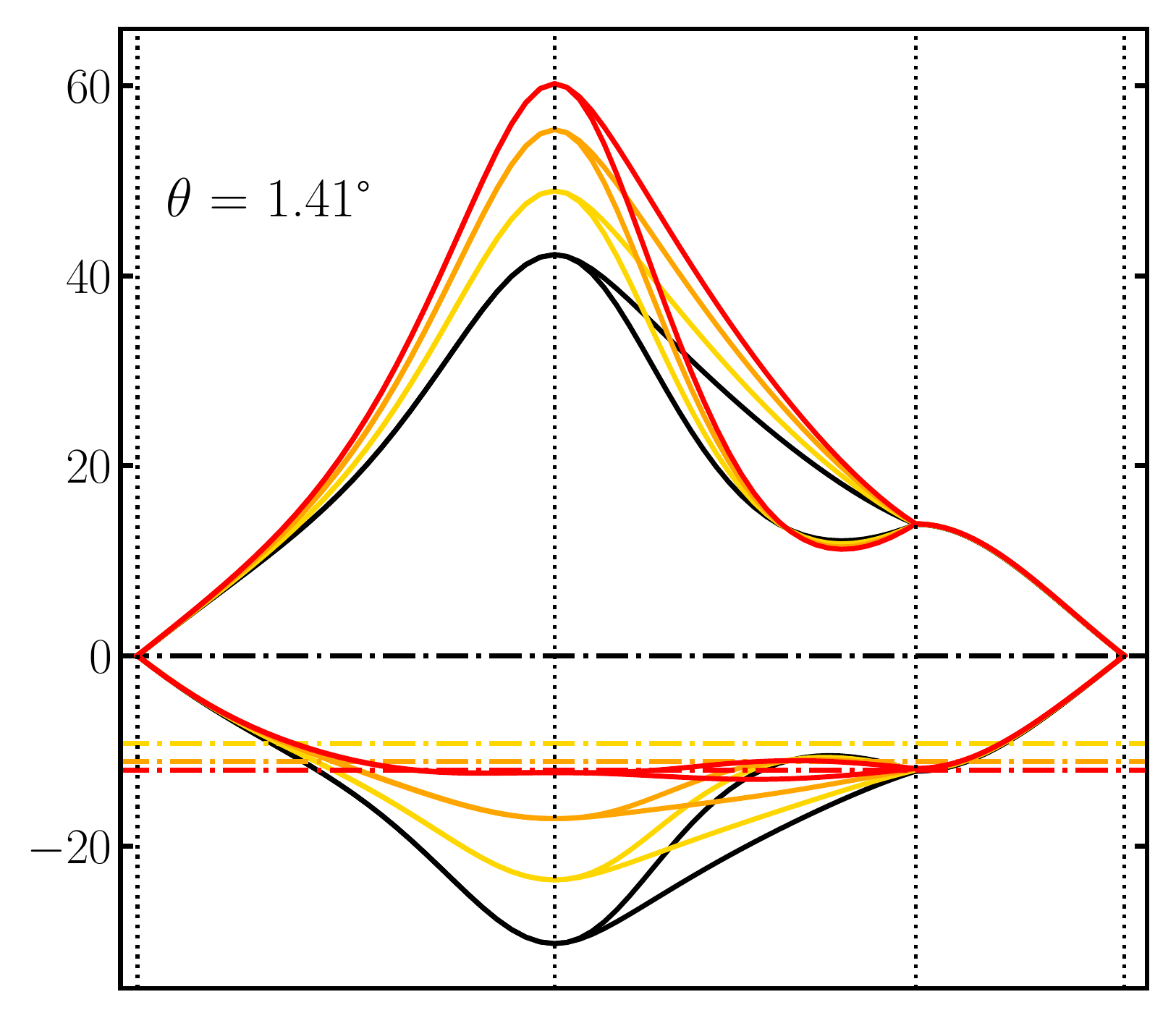}
\end{subfigure}
\begin{subfigure}{0.3\textwidth}
  \includegraphics[width=1\linewidth]{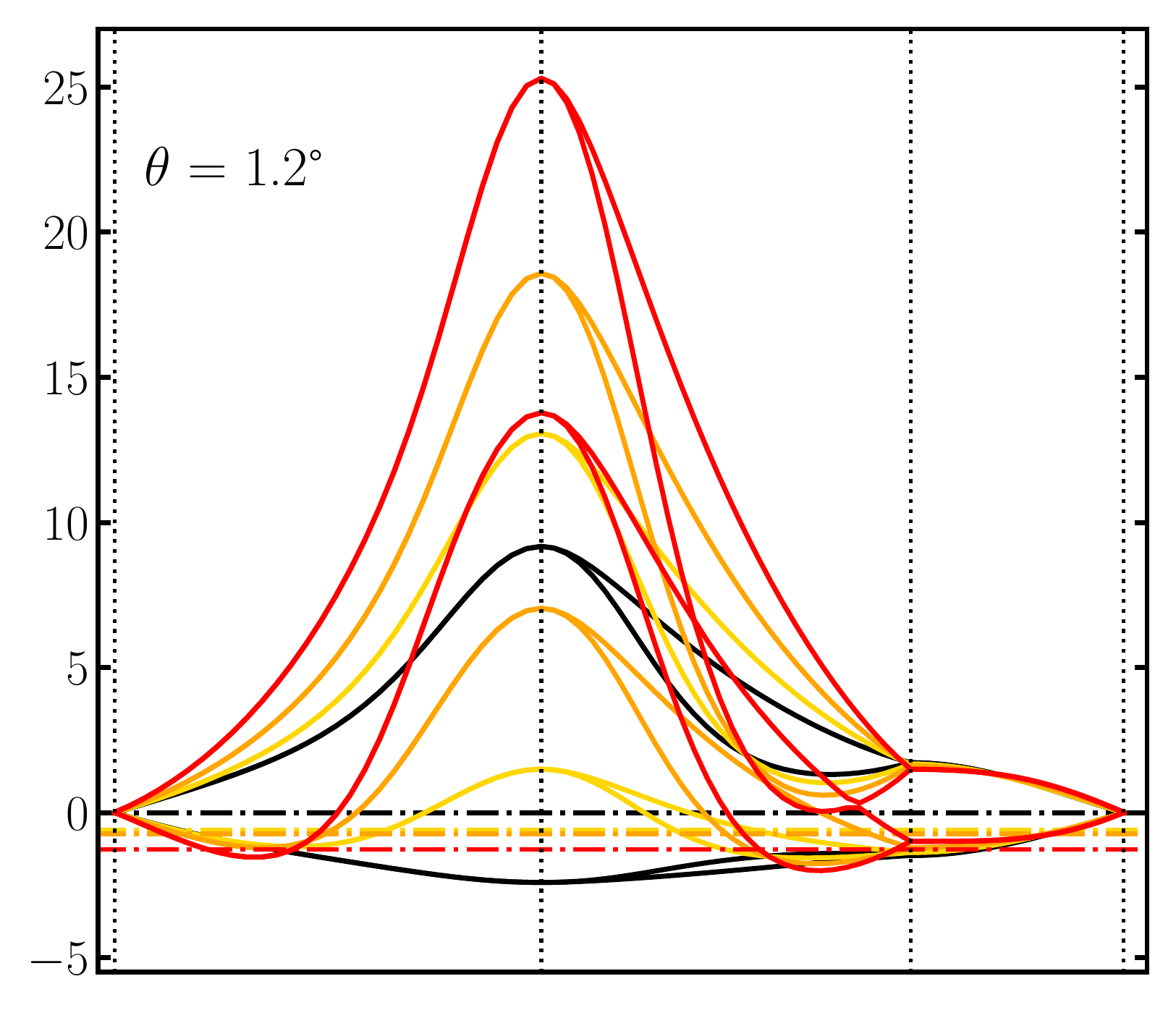}
\end{subfigure}
\begin{subfigure}{0.325\textwidth}
  \includegraphics[width=1\linewidth]{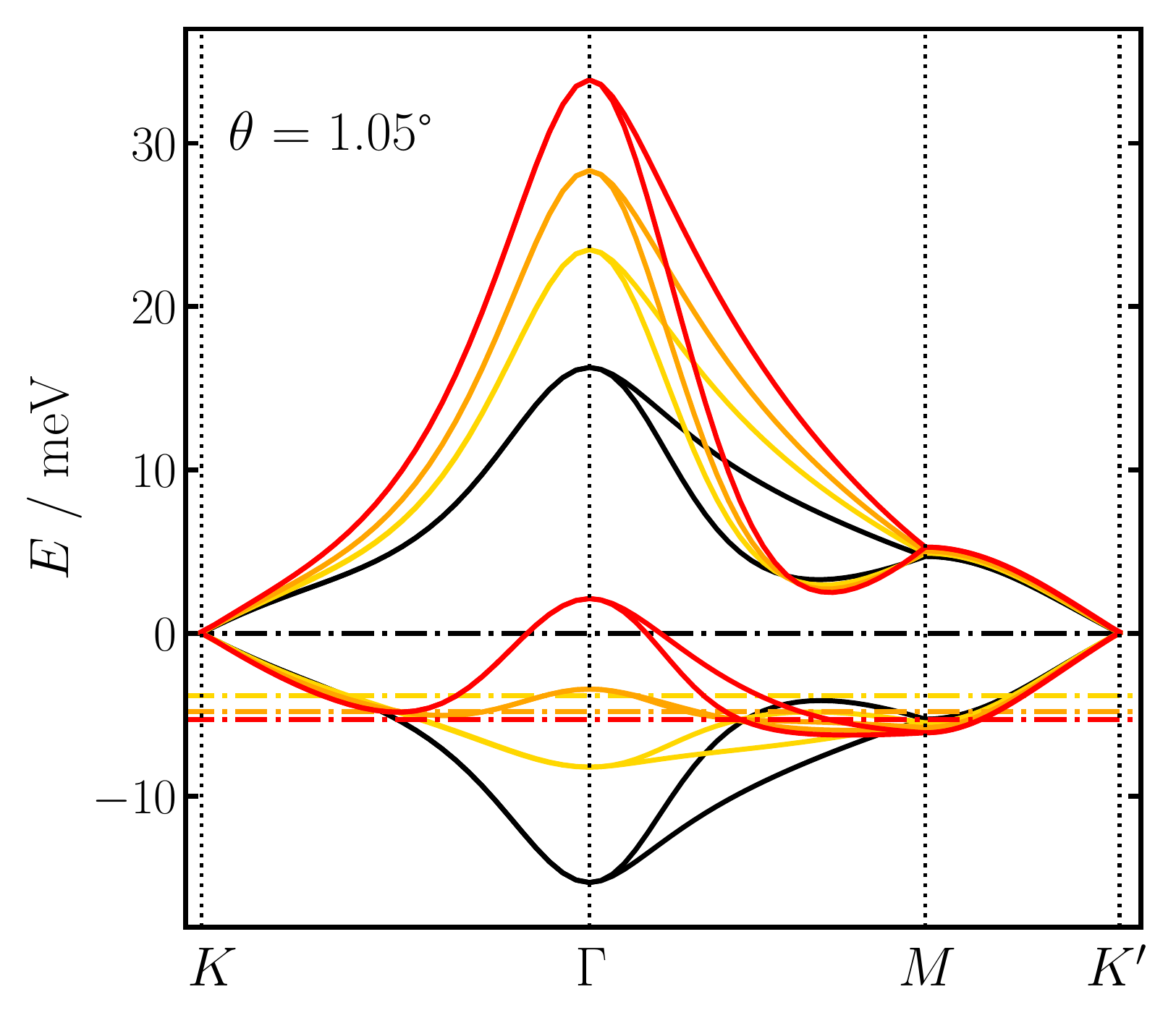}
\end{subfigure}
\begin{subfigure}{0.3\textwidth}
  \includegraphics[width=1\linewidth]{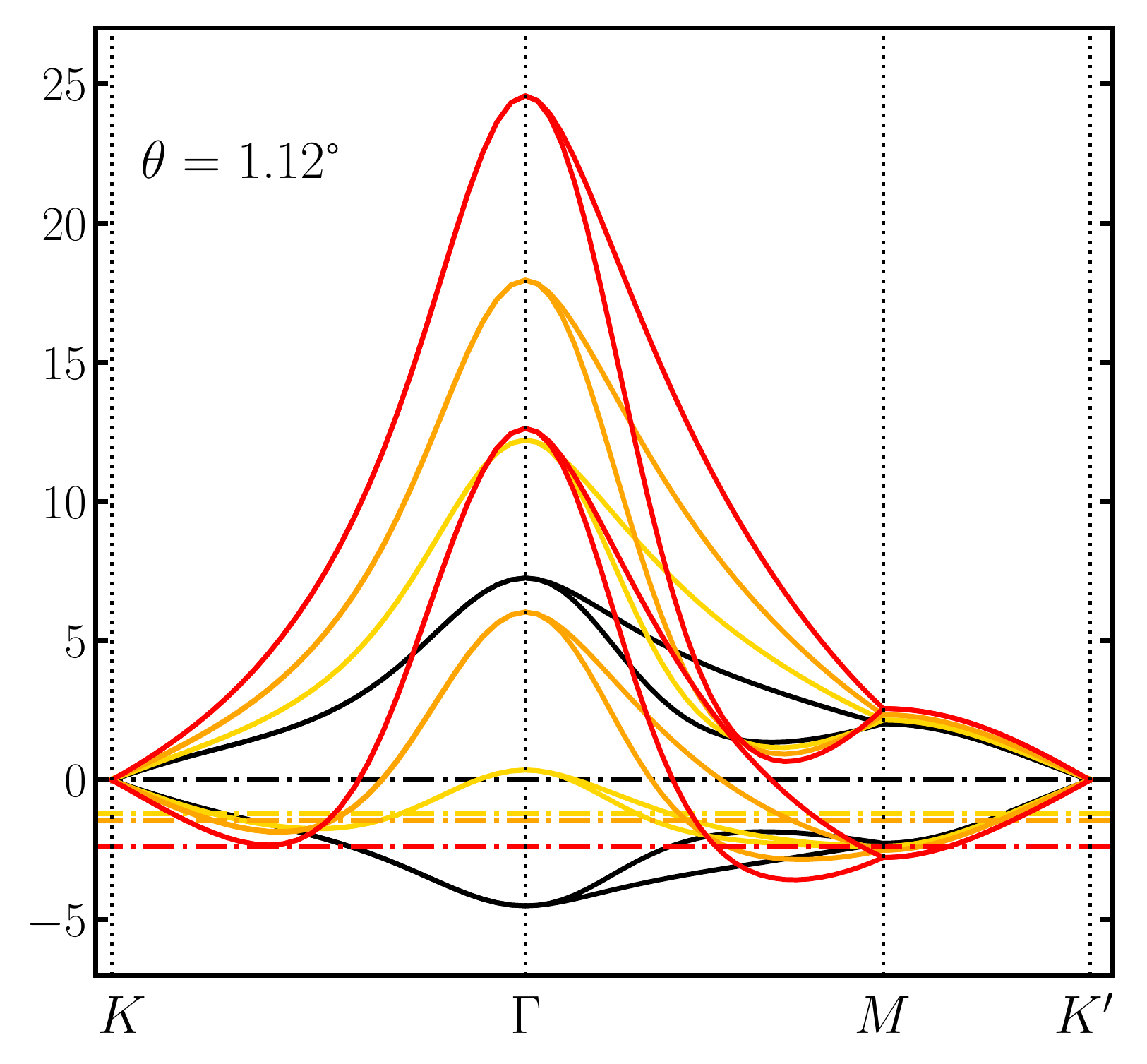}
\end{subfigure}
\begin{subfigure}{0.3\textwidth}
  \includegraphics[width=1\linewidth]{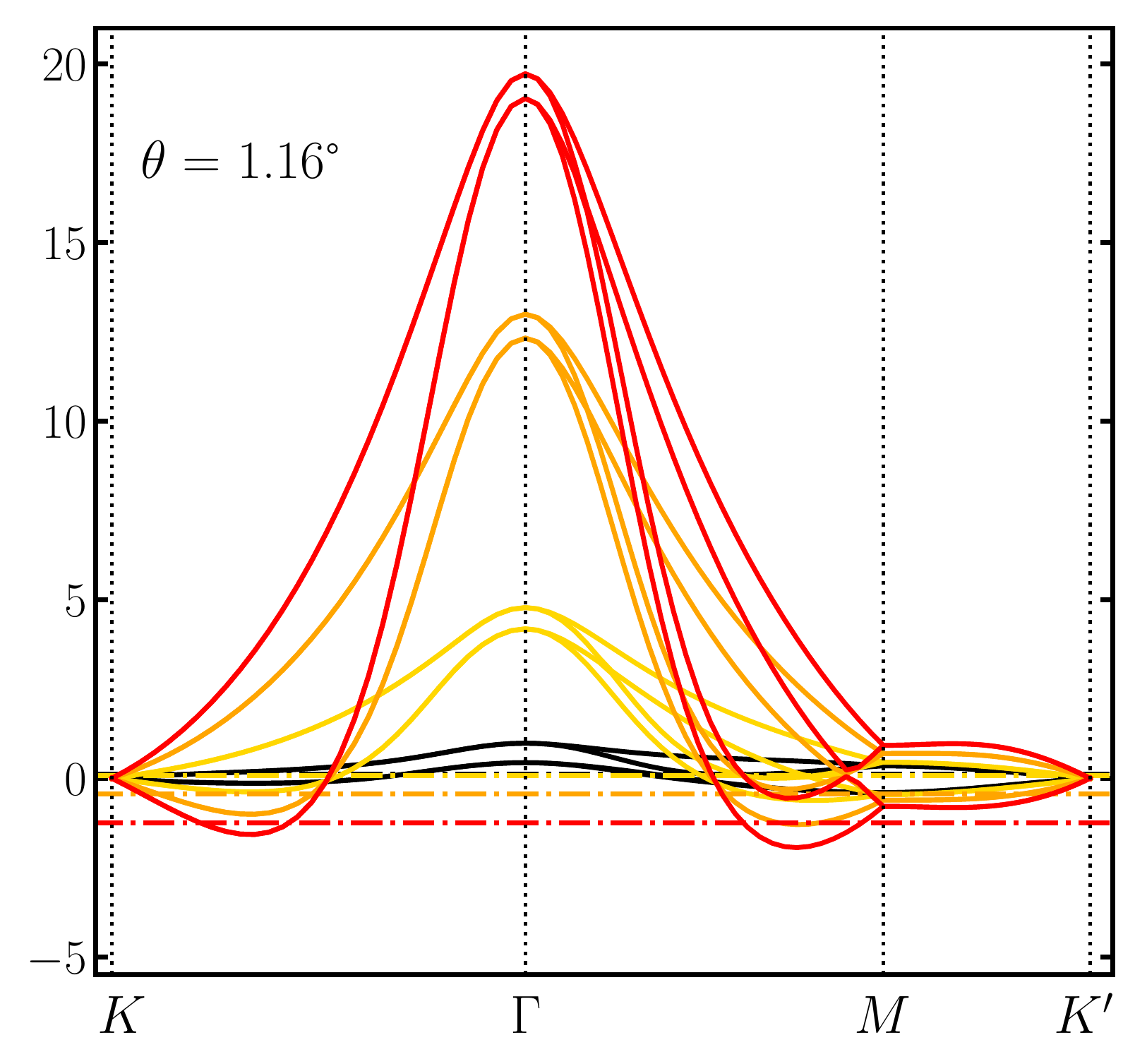}
\end{subfigure}
\caption{Atomistic Hartree band structures of twisted bilayer graphene for various twist angles $\theta$ and doping levels $\nu$, assuming dielectric screening with $\epsilon_\mathrm{bg}=1$. Band structures of electron-doped (hole-doped) tBLG are shown in the upper (lower) two rows; the undoped case (black line) is shown in all panels. The Fermi level is denoted by horizontal dash-dotted lines. For clarity, the energy at the $K$-point is used as reference in all graphs and only the four flat bands near the Fermi level are shown. Note that the width of the flat band manifold and therefore also the scale of the vertical axis depends sensitively on the twist angle, as shown in Fig.~\ref{fig:BW}. The width of the flat band manifold is smallest at $\theta=1.16\degree$.}
\label{fig:BS}
\end{figure*}

\begin{figure*}[t!]
\centering
\begin{subfigure}{0.45\textwidth}
  \includegraphics[width=1\linewidth]{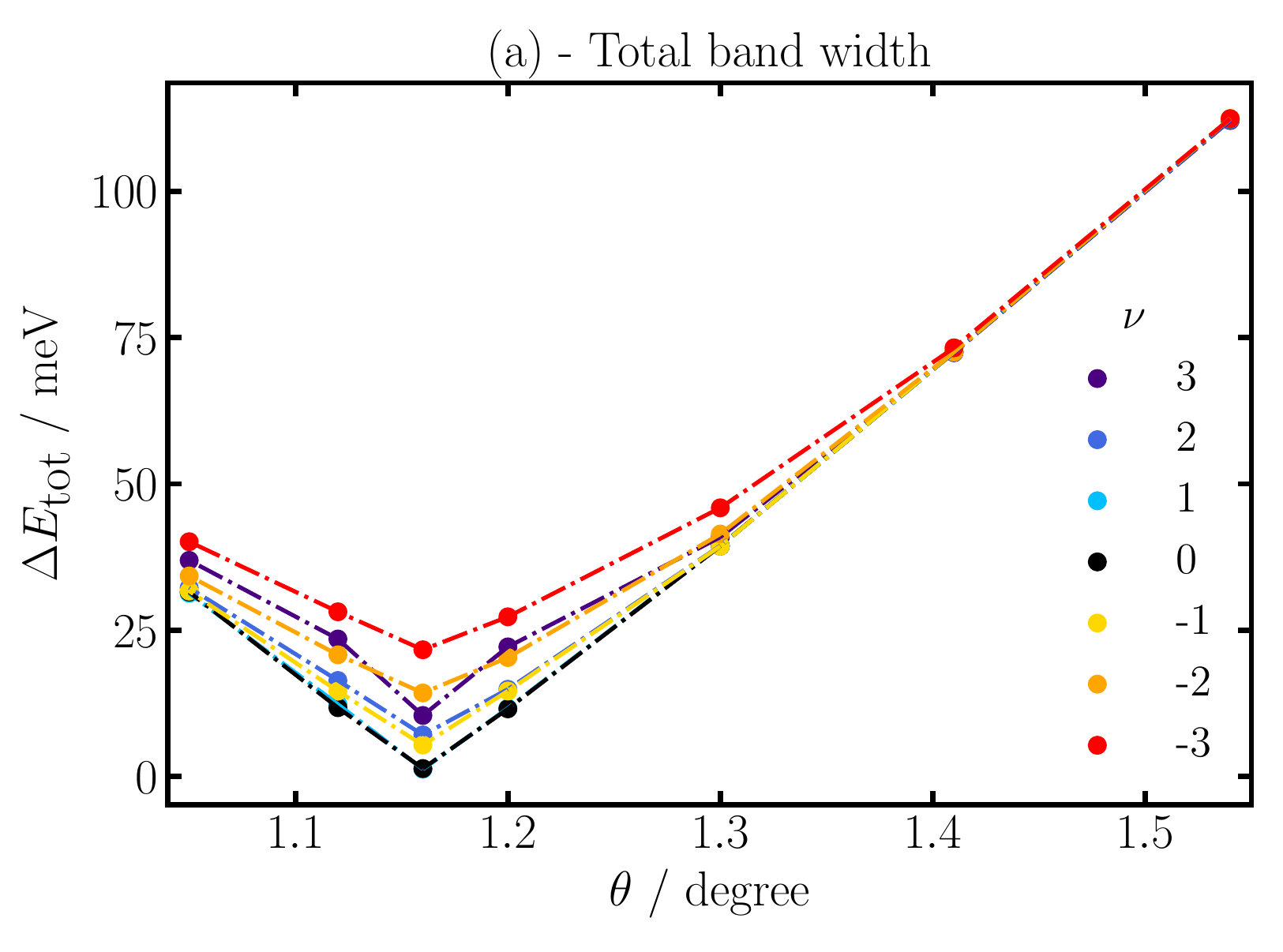}
\end{subfigure}
\begin{subfigure}{0.45\textwidth}
  \includegraphics[width=1\linewidth]{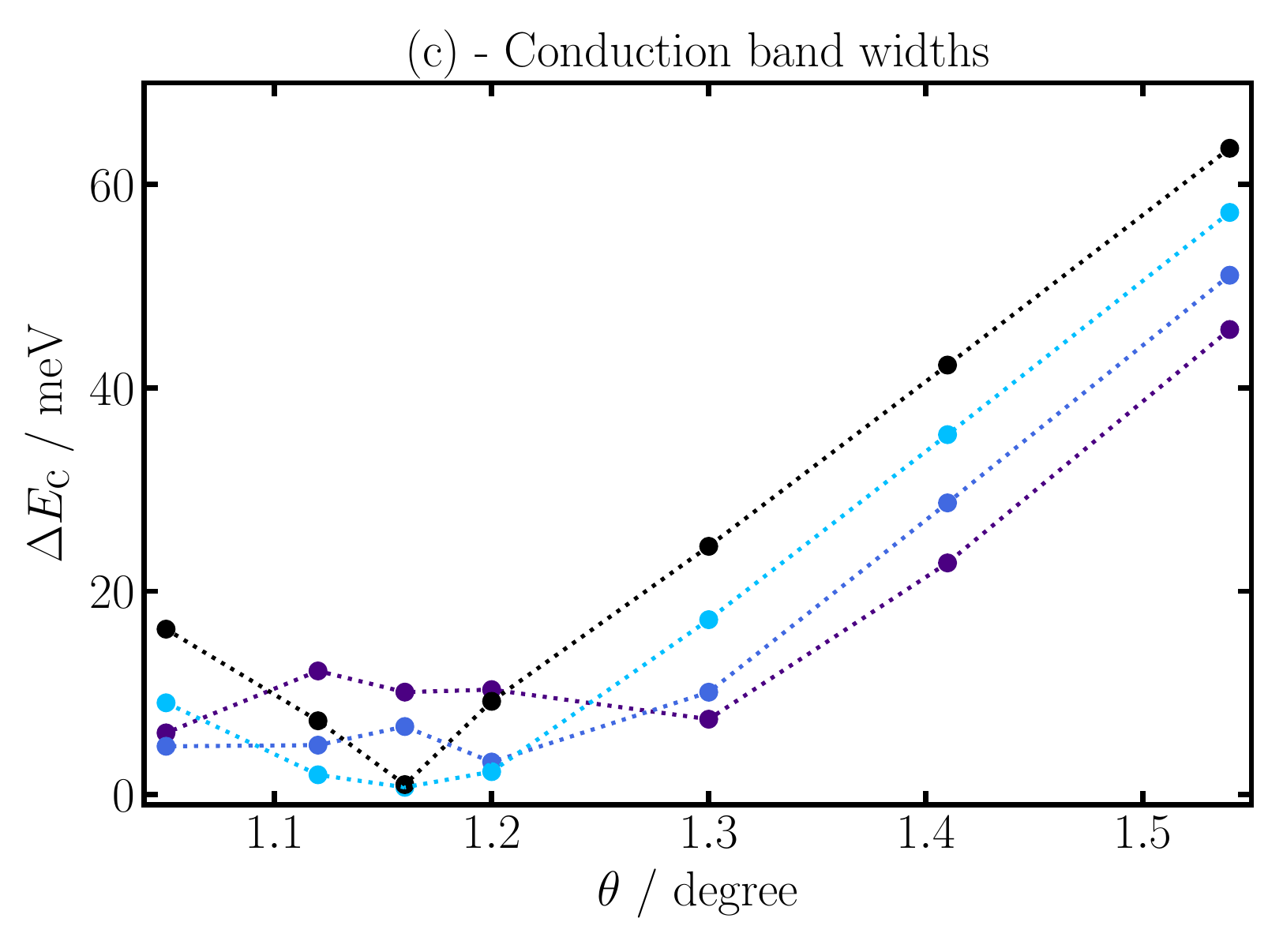}
\end{subfigure}
\begin{subfigure}{0.45\textwidth}
  \includegraphics[width=1\linewidth]{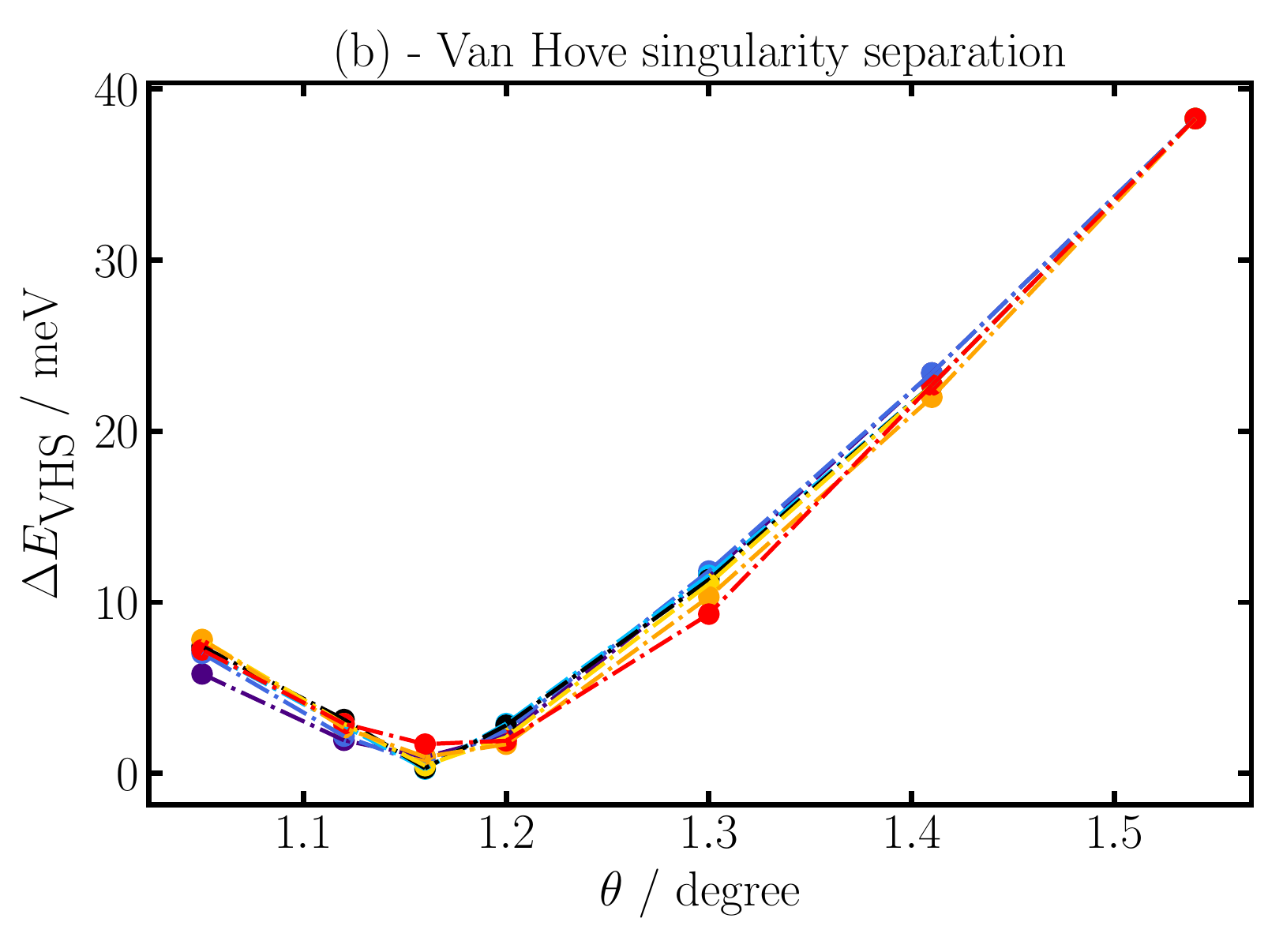}
\end{subfigure}
\begin{subfigure}{0.45\textwidth}
  \includegraphics[width=1\linewidth]{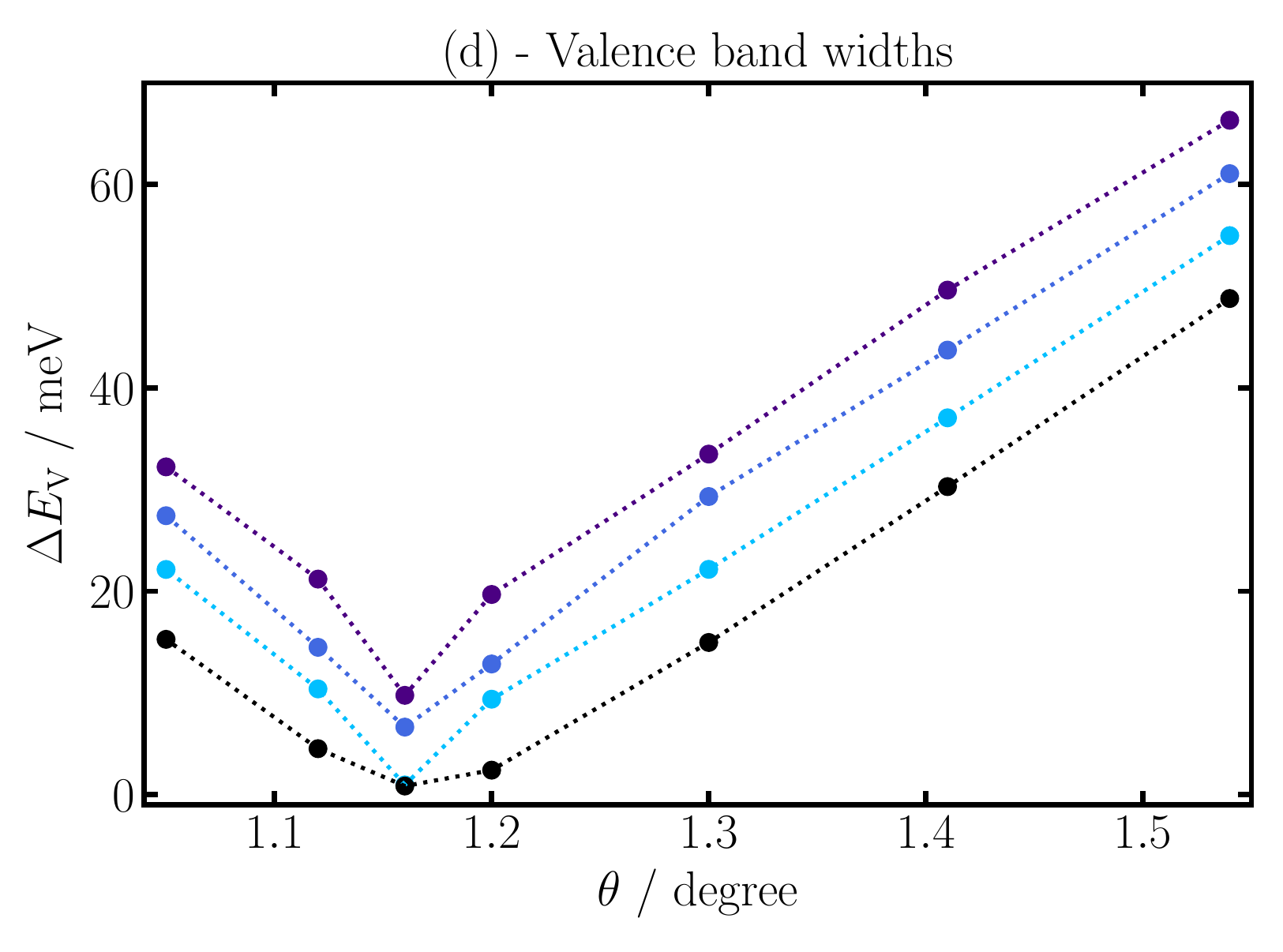}
\end{subfigure}
\caption{(a) Width $\Delta E_{\text{tot}}$ of the flat band manifold as a function of twist angle for different doping levels. (b) Energy separation $\Delta E_{\text{VHS}}$ of the valence and conduction van Hove singularities (VHS) as function of twist angle for different doping levels. (c) Width $\Delta E_\mathrm{c}$ of the flat conduction bands as function of twist angle for different doping levels $\nu\ge 0$. (d) Width $\Delta E_\mathrm{v}$ of the flat valence bands as function of twist angle for different doping levels $\nu\ge 0$. In all cases we assume dielectric screening with $\epsilon_\mathrm{bg}=1$.}
\label{fig:BW}
\end{figure*}

\section{Results and Discussion}

\subsection{Band structure}

Figure~\ref{fig:BS} shows the band structures of tBLG suspended in air ($\epsilon_\mathrm{bg}=1$) from Hartree theory at six twist angles between $\theta=1.54\degree$ and $\theta=1.05\degree$ (below, we show that a dielectric substrate only leads to small changes in the Hartree band structures). Only the four flat bands closest to the Fermi energy are shown. We refer to the lower two of the flat bands as the flat valence bands and the upper two as the flat conduction bands. These four bands are separated from all other bands by energy gaps that result from the atomic corrugation of tBLG~\cite{MLWO,PHD_1,CRAC,KDP,LREBM,EDS}. The width of the flat band manifold is smallest at $\theta=1.16\degree$ and we refer to this twist angle as the magic angle. 

We first discuss the band structures of undoped tBLG, corresponding to $\nu=0$ (black curves in Fig.~\ref{fig:BS}). The band structures at all twist angles except the magic angle are semi-metallic and feature linear bands at the $K$ and $K'$ points. As the magic angle is approached, the total width of the flat band manifold decreases rapidly, see Fig.~\ref{fig:BW}(a). Interestingly, at charge neutrality, the valence band widths are always smaller than the conduction band widths, see Figs.~\ref{fig:BW}(c) and (d).

At the magic angle, the band structure of undoped tBLG is qualitatively different as compared to the other twist angles~\cite{KDP,EPC}. In particular, the two valence bands at $\Gamma$ are pushed up and are now higher in energy than the states at $K$ and $K'$. As a consequence, at this level of theory, tBLG is metallic at the magic angle even without doping with additional carriers. 

The Hartree band structures of undoped tBLG are very similar to the non-interacting tight-binding band structures (see Supplementary Materials for a comparison). This can be understood by analyzing the charge density and the corresponding Hartree potential. Without doping the charge density oscillates on the atomic scale, but each region of the moir\'e unit cell is approximately charge neutral (when the atomic oscillations are averaged over a region)~\cite{H_AM} resulting in a small Hartree potential [Eq.~\eqref{eq:VH}], as we shall discuss further later.

Figure~\ref{fig:BS} also shows Hartree band structures for electron-doped (upper two rows) and hole-doped (lower two rows) tBLG. In agreement with previous Hartree calculations~\cite{H_CM,H_AM,EE}, we observe that doping results in significant changes in the band structures. In contrast, the tight-binding band structures that are widely used to understand the electronic properties of doped tBLG do not change upon doping. Focusing first on the largest twist angle considered, $\theta=1.54\degree$, electron doping (corresponding to $\nu=1$, 2 or 3) flattens the conduction bands, while the valence bands become more dispersive. Figure~\ref{fig:BW}(c) shows that the conduction band width decreases by approximately $5$~meV for each added electron. However, the valence band width increases by the same amount [Fig.~\ref{fig:BW}(d)] and the total band width of the flat band manifold remains constant at this twist angle, as shown in Fig.~\ref{fig:BW}(a). For hole doping ($\nu=-1$, $-2$ and $-3$), the situation is similar but the valence bands flatten and the conduction bands become more dispersive. 

To understand why electron-electron interactions are more relevant for the doped system, we analyze again the charge density and the corresponding Hartree potential (the explanation here follows that outlined by Rademaker \textit{et al.} in Ref.~\cite{H_AM}). As the local density of states is larger in the AA regions than in the AB/BA regions, additional carriers (both electrons and holes) preferentially localize in the AA regions~\cite{H_AM}. This creates a highly non-uniform charge distribution, which gives rise a strong Hartree potential~\cite{H_AM}. Fig.~\ref{fig:VH}(a) shows that $\Delta V_\mathrm{H}$ (the difference between the Hartree potential in the centers of the AA and AB regions) increases by approximately $30$~meV for each added electron. States near the $K$ and $K'$ points are localized in the AA regions and are pushed up in energy relative to the states at $\Gamma$ (which have a ring-like shape surrounding the AA regions) for electron-doped systems~\cite{H_AM}. In contrast, the $K$/$K'$ states are pushed down in energy relative to the $\Gamma$-states for hole-doped systems~\cite{H_AM}. 

For smaller twist angles, doping induces even more significant changes in the band structure. At $\theta=1.41\degree$, the valence bands are almost completely flat between $\Gamma$ and $M$ for $\nu=-3$. In contrast, the flattening of the conduction bands upon electron doping is not quite as pronounced at this twist angle. For $\theta=1.2\degree$, the $\Gamma$-states have moved past the $K$/$K'$-states so that the curvature of the conduction band at $\Gamma$ changes sign at all doping levels (both electron and hole doping) except $\nu=1$. For this doping level, the conduction band is very flat in the vicinity of the $\Gamma$-point. Interestingly, for $\nu=2$ the width of the conduction bands exhibits a local minimum at $\theta=1.2\degree$, see Fig.~\ref{fig:BW}(c), and is even smaller than at the magic angle (defined as the twist angle that exhibits the smallest total band width of the flat band manifold, $\theta=1.16\degree$). Similarly, for $\nu=3$ the width of the conduction bands exhibits a local minimum at $\theta=1.3\degree$. This suggests that long-ranged Coulomb interactions between electrons can modify the twist angle at which electron correlation phenomena are strongest and that this may not necessarily be at the magic angle.

These qualitative changes in the band structures of doped tBLG close to the magic angle can be understood by analyzing the twist angle dependence of the Hartree potential. Fig.~\ref{fig:VH}(a) shows that $\Delta V_\mathrm{H}$ only depends weakly on the twist angle. In contrast, the band widths decrease rapidly as the magic angle is approached and therefore the importance of long-ranged electron-electron increases strongly. 

At the magic angle ($\theta=1.16\degree$), the band structures of hole-doped tBLG ($\nu=-1$, $-2$ and $-3$) look qualitatively similar to the undoped band structure, but with a significantly larger band width. For example, for $\nu=-2$ we find a band width of 13~meV (compared to 1~meV for the undoped system). For electron-doped systems, the conduction bands `invert' such that both the valence and conduction bands at $\Gamma$ are at lower energies than the states at $K$ and $K'$. 

For twist angles smaller than the magic angle, the band structures of doped tBLG are quite similar to those of twist angles larger than the magic angle. In particular, the band structures at $\theta=1.12\degree$ correspond closely to those of $\theta=1.2\degree$ (both differ from the magic angle by $0.04\degree$) and the band structures of $\theta=1.05\degree$ are similar to those of $\theta=1.3\degree$ (which differ from the magic angle by $0.11\degree$ and $0.14\degree$, respectively). 

\begin{figure*}[t!]
\centering
\begin{subfigure}{0.325\textwidth}
  \includegraphics[width=1\linewidth]{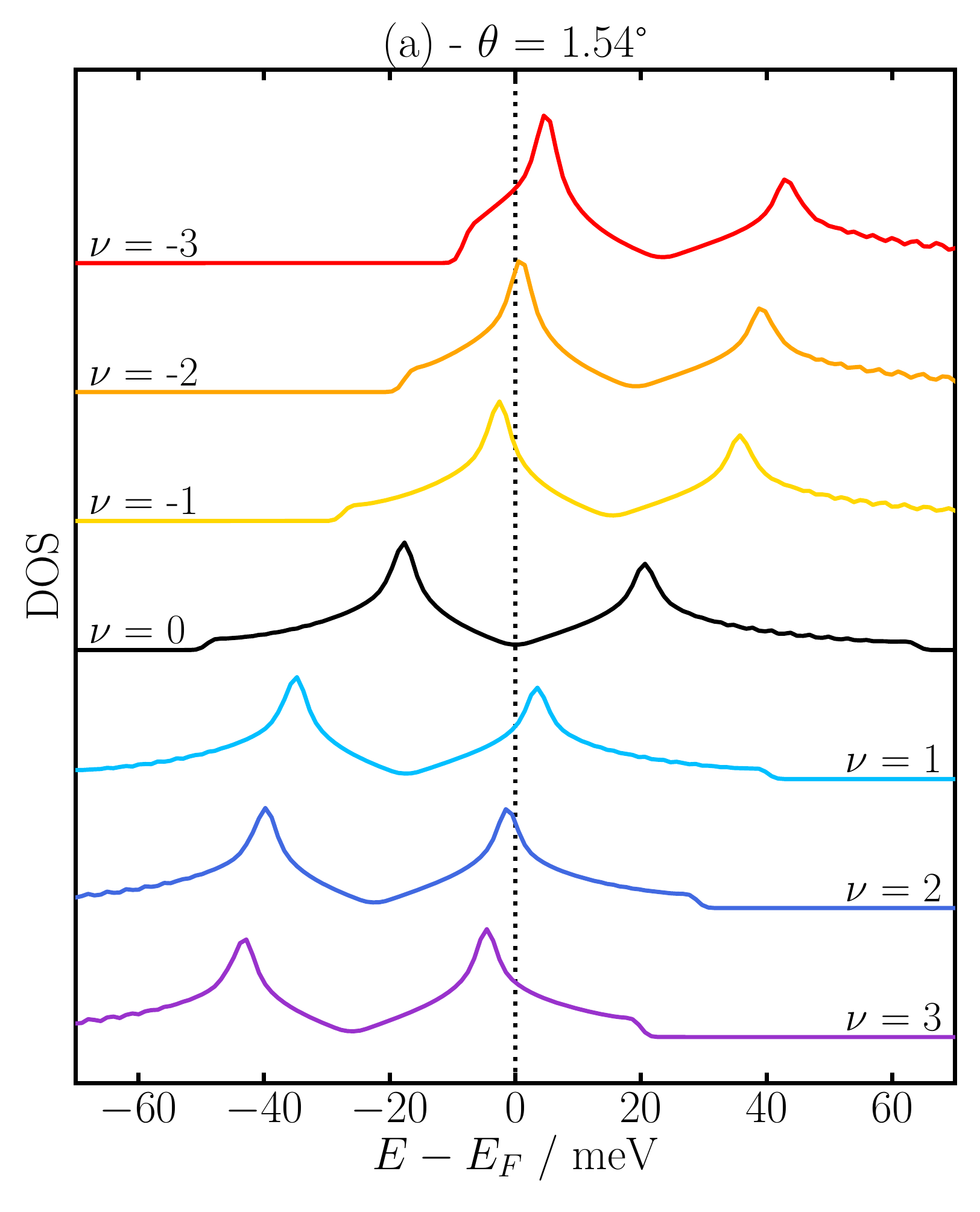}
\end{subfigure}
\begin{subfigure}{0.325\textwidth}
  \includegraphics[width=1\linewidth]{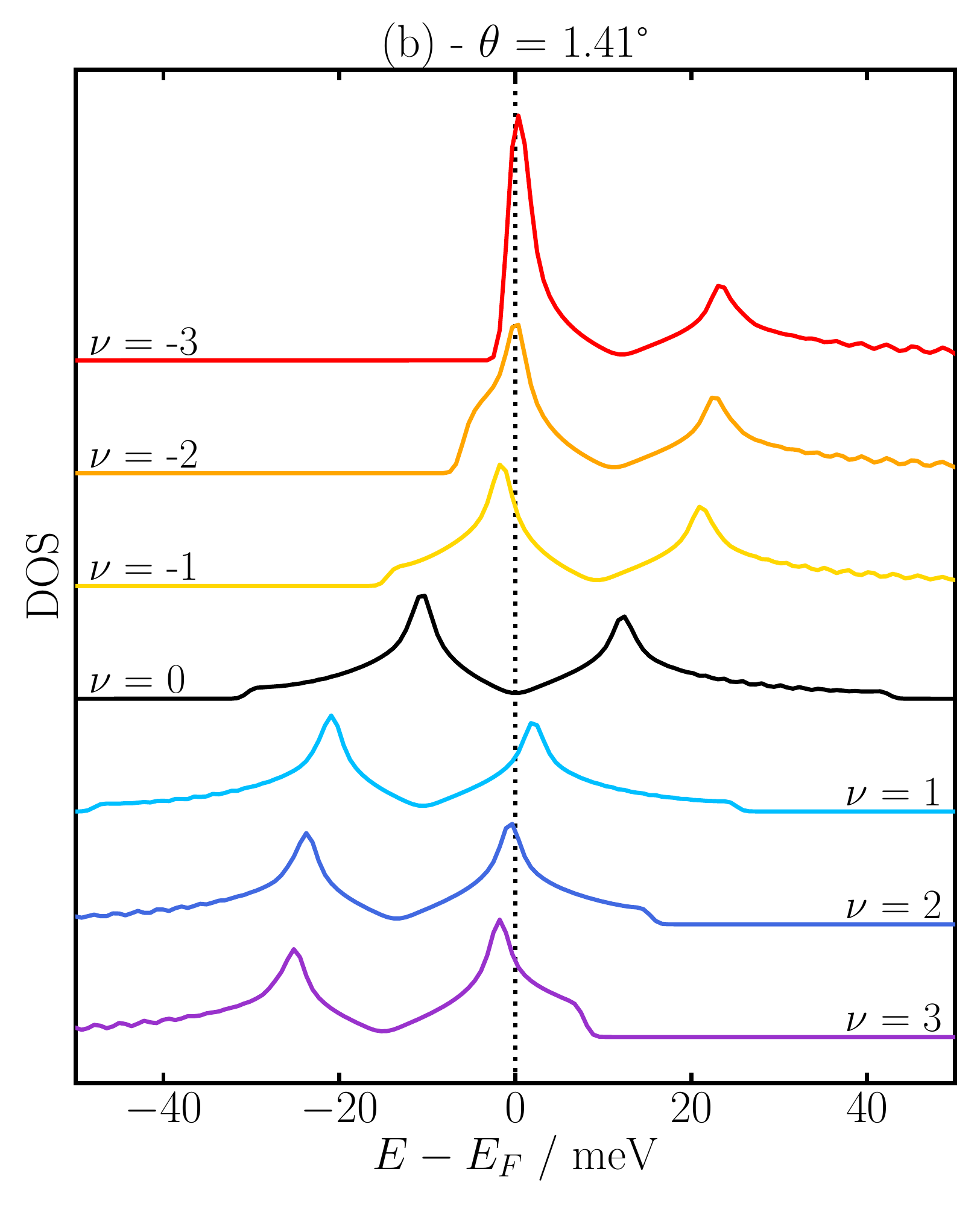}
\end{subfigure}
\begin{subfigure}{0.325\textwidth}
  \includegraphics[width=1\linewidth]{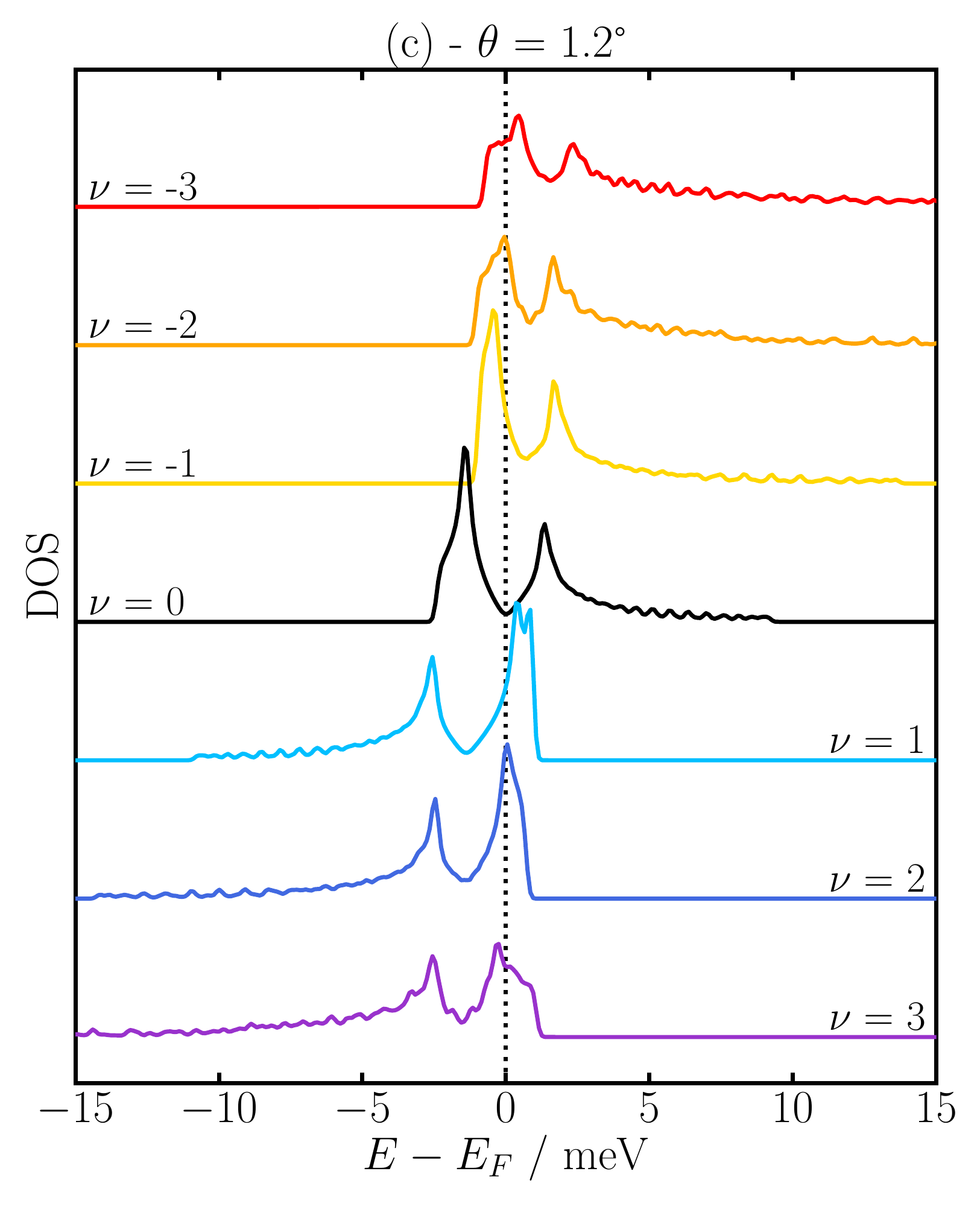}
\end{subfigure}
\caption{Doping-dependent densities of states (DOS) of twisted bilayer graphene (suspended in air) for three twist angles near the magic angle. The dotted vertical line denotes the Fermi level. Additional results for other twist angles are shown in the Supplementary Materials.}
\label{fig:DOS}
\centering
\begin{subfigure}{0.325\textwidth}
  \includegraphics[width=1\linewidth]{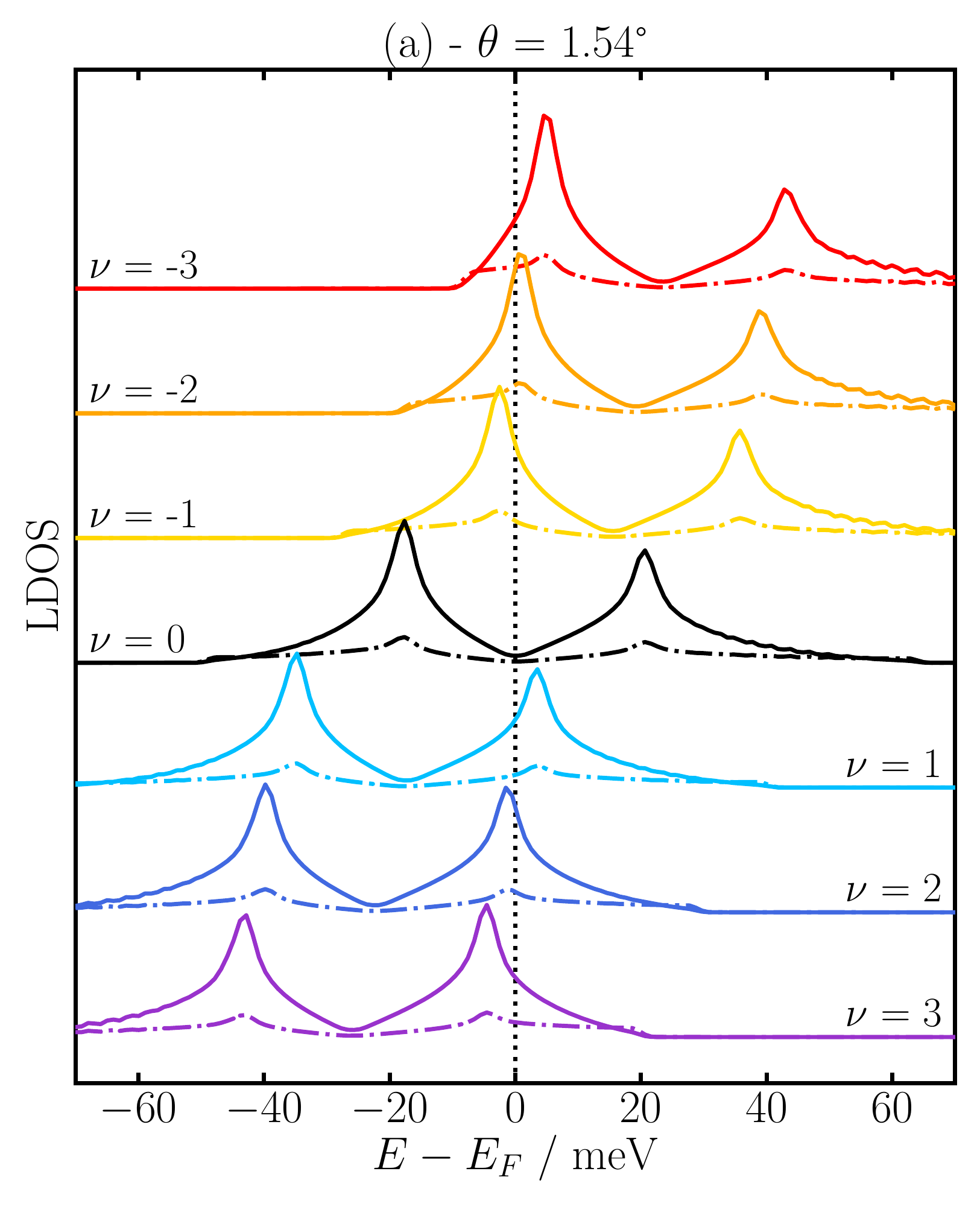}
\end{subfigure}
\begin{subfigure}{0.325\textwidth}
  \includegraphics[width=1\linewidth]{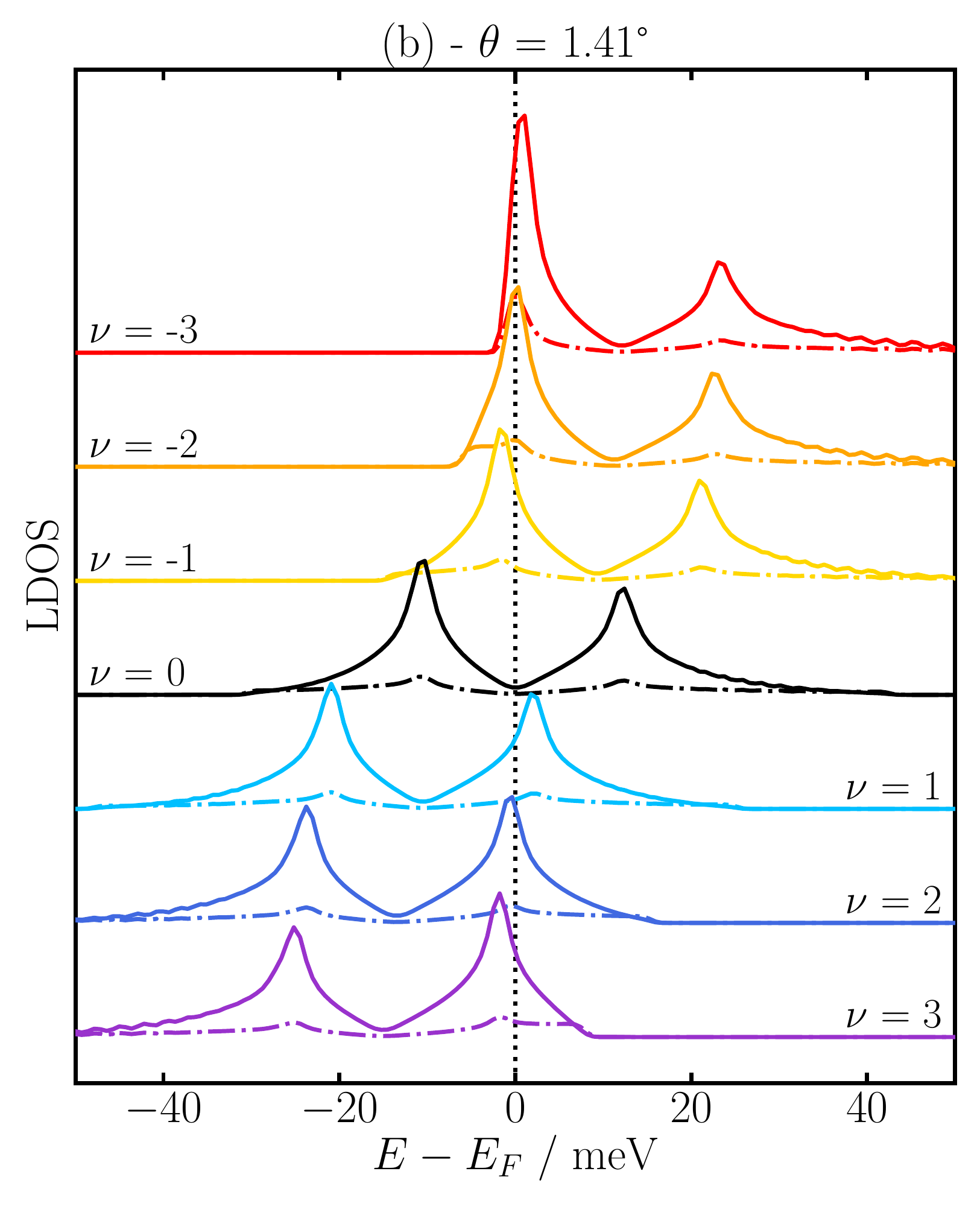}
\end{subfigure}
\begin{subfigure}{0.325\textwidth}
  \includegraphics[width=1\linewidth]{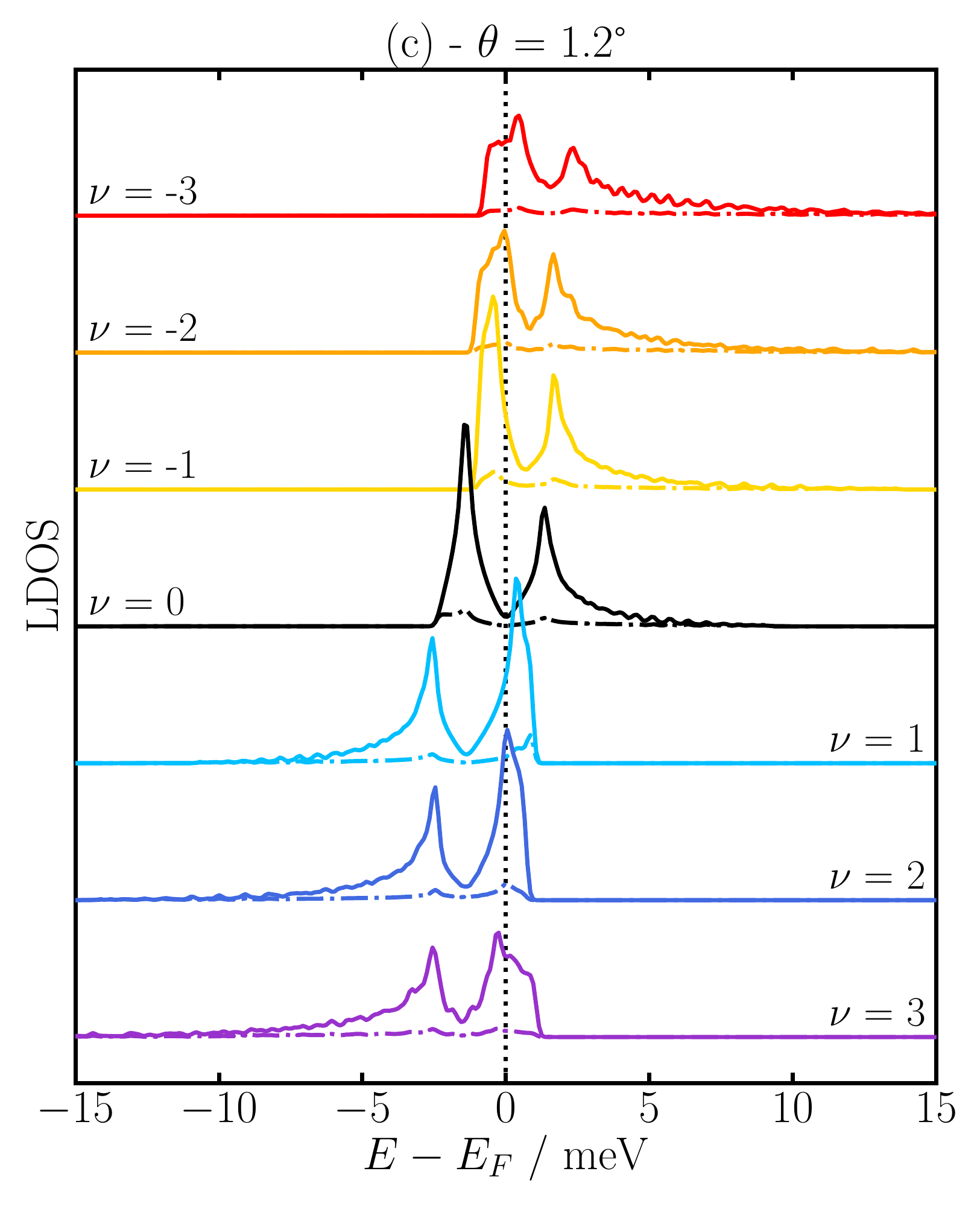}
\end{subfigure}
\caption{Doping-dependent local densities of states (LDOS) in the AA (solid curves) and AB (dash-dotted curves) regions of twisted bilayer graphene (suspended in air) for three twist angles near the magic angle. The dotted vertical line denotes the Fermi level. Additional results for other twist angles are shown in the Supplementary Materials.}
\label{fig:LDOS}
\end{figure*}

\subsection{DOS and LDOS}

Figures~\ref{fig:DOS} and \ref{fig:LDOS} show the DOS and LDOS from Hartree theory for three twist angles: $\theta=1.54\degree$ [panel (a)], $\theta=1.41\degree$ [panel (b)] and $\theta=1.2\degree$ [panel (c)]. The LDOS is shown both for the AA (solid lines) and AB (dash-dotted lines) regions, averaged over a region around the centre of the respective region (as discussed in the Methods section). When the tunnelling matrix elements are constant (which is likely a good approximation for the flat bands of tBLG), the LDOS is proportional to the measured tunnelling spectrum and thus directly accessible in experiments. Several STS studies of tBLG have been reported recently~\cite{NAT_SS,NAT_MEI,NAT_CO} and we will discuss the similarities and differences of our calculations with these experimental measurements. Below, we analyse each of the three twist angles in turn; the results for the other twist angles can be found in the Supplementary Materials. 

For $\theta=1.54\degree$, the DOS exhibits two pronounced peaks at all doping levels. At zero doping, these van Hove singularities (VHS) are located at $\pm20$~meV on both sides of the Fermi energy (so their energy separation is approximately $40$~meV). Comparing the DOS to the LDOS, we find that the dominant contribution to the DOS derives from the AA regions~\cite{LDE}. The LDOS in the AB regions also exhibits small peaks in the vicinity of the VHS. The valence band VHS is somewhat larger than the conduction band one because the valence bands are flatter than the conduction bands, see Fig.~\ref{fig:BS}. These findings are in agreement with several recent experimental STS measurements~\cite{NAT_SS,NAT_MEI,NAT_CO}. Note, however, that our values for the energy difference between valence and conduction VHS are smaller (for the same twist angle) than the experimental results. In Ref.~\citenum{NAT_MEI}, this was attributed to the use of DFT-derived tight-binding parameters for the interlayer hopping which are about 20 percent smaller than more accurate GW values. 

Upon electron doping, the conduction VHS becomes larger while the valence VHS becomes smaller. This is a consequence of the doping-induced band flattening of the conduction bands, while the valence bands become more dispersive, see Fig.~\ref{fig:BS}. In contrast, hole doping increases the valence VHS while the conduction VHS becomes smaller. Again, these findings are in agreement with several recent experimental measurements and cannot be explained by tight-binding theory. Note that at this twist angle the Fermi level of the doped system is not pinned at the VHS. 

At $\theta=1.41\degree$, the separation between the VHS is reduced to $30$~meV. Upon hole doping, the difference between valence and conduction band VHS is much clearer than at $1.54\degree$. This is caused by the strong distortion of the doped valence bands resulting in extremely flat valence bands throughout large regions of the Brillouin zone, see Fig.~\ref{fig:BS} (recall that the distortion of the valence bands is always more pronounced that that of the conduction bands). For $\nu=-2$ and $\nu=-3$, we observe that the Fermi level is pinned at the valence VHS. This Fermi level pinning has also been reported in several experimental STS studies and is a consequence of electron-electron interaction induced changes in the band structure. The LDOS in the AA region is again very similar to the DOS. However, we find that the valence peak of the LDOS in the AB regions grows upon hole doping (see SM for further details). This is because the wave functions of the flat valence bands are partly localized in the AB regions (in particular, the valence states near $\Gamma$). This prediction can be tested by STM measurements and would provide direct evidence of the doping-induced band flattening in Hartree theory. Figure~\ref{fig:BW}(b) shows that the separation of the VHS is reduced by hole doping for twist angles larger than the magic angle and increased by electron doping. The opposite trend is observed for twist angles smaller than the magic angle. While this is in qualitative agreement with some experimental measurements, the absolute magnitude of the change in VHS separation is typically smaller than in experiments~\cite{ECNMA,NAT_SS,NAT_MEI,NAT_CO,TMBSC,OVH,VHSDTBLG,ADVHS,SvHSstblg,PWCS}.

Besides Fermi level pinning, the enhancement of the DOS at the Fermi level due to the doping-induced flattening of the partially occupied bands is also relevant for understanding broken-symmetry phases, such as correlated insulator or superconducting states~\cite{SSCmfb}. In particular, the values of the transition temperatures to these states are usually very sensitive to the DOS at the Fermi energy~\cite{SSCmfb}, $\textrm{DOS}(E_\textrm{F})$. For example, the superconducting critical transition temperature is given by $T_\mathrm{c} \propto \exp{(-1/[\textrm{DOS}(E_{\textrm{F}})V])}$ with $V$ describing the coupling strength of the electrons to the superconducting glue (e.g., phonons or spin waves). The doping-induced increase of the DOS at the Fermi level should therefore result in a dramatic increase of the critical temperature. Again, this effect is not captured by tight-binding theory.

At $\theta=1.2\degree$, very close to the magic angle, the VHS separation is only $5$~meV in the undoped system and the valence VHS is much larger than the conduction VHS. Fermi level pinning is observed both for electron and hole doping. In the DOS, the shape of the VHS of the partially filled band is highly asymmetric. In particular, the leading edge of the peak (i.e., the side of the peak facing towards the other VHS) rises more sharply than the trailing edge (i.e., the side facing away from the other VHS). Interestingly, we also observe a double peak in the conduction VHS at $\nu=1$. The second peak is caused by a peak of the LDOS in the AB regions which does not coincide with the main peak of the LDOS in the AA regions. Again, this double peak structure is caused by the electron-electron interaction induced distortion of the conduction band near $\Gamma$. Fig.~\ref{fig:BS} shows that the conduction bands are extremely flat near $\Gamma$, but have a slightly higher energy than the states at $M$ which give rise to the main peak of the VHS.

\begin{figure}[h]
\centering
\begin{subfigure}{0.45\textwidth}
\includegraphics[width=1\linewidth]{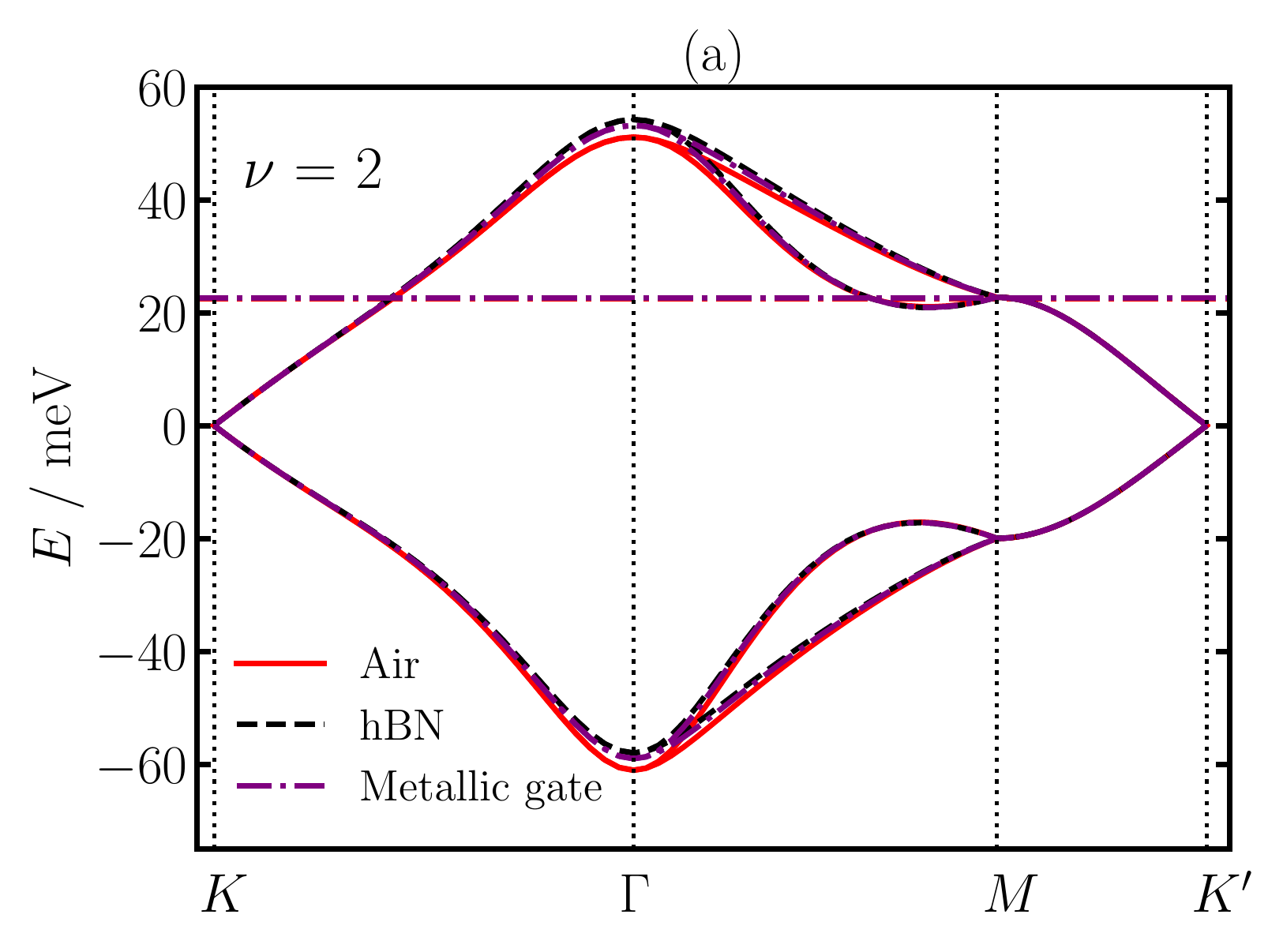}
\end{subfigure}
\begin{subfigure}{0.45\textwidth}
  \includegraphics[width=1\linewidth]{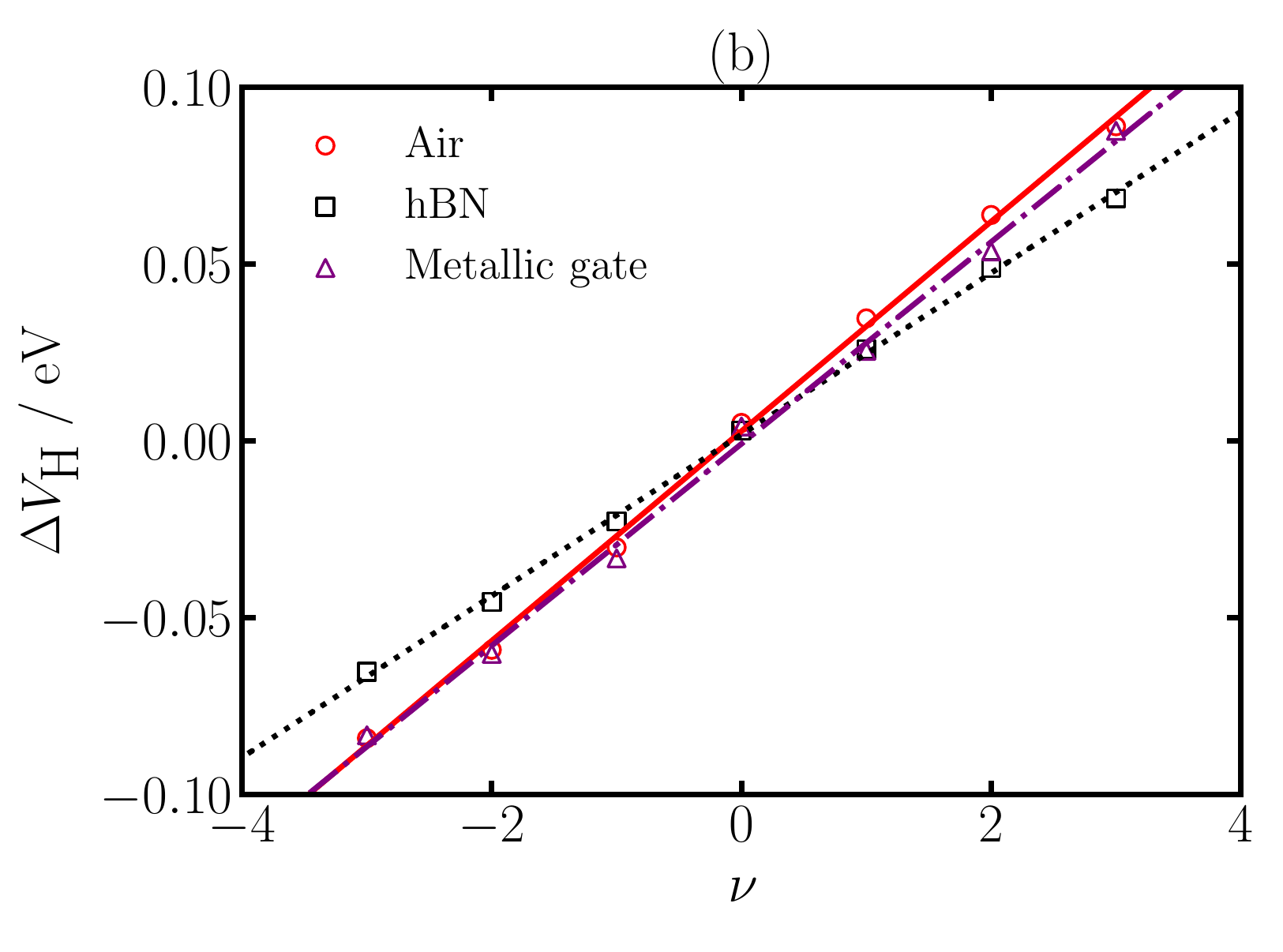}
\end{subfigure}
\caption{(a) Band structure of electron-doped ($\nu=2$) twisted bilayer graphene suspended in air ($\epsilon_\mathrm{bg}=1$; red solid lines), encapsulated by hBN ($\epsilon_\mathrm{bg}=3.9$; black dashed lines) and encapsulated by hBN with metallic gates at a distance of $10~\textrm{\AA}$ (purple dash-dotted line). (b) Corresponding values of $\Delta V_\mathrm{H}$ defined as the difference of the Hartree potential energy in the centers of the AA and the AB regions. The twist angle is $\theta=1.54\degree$.}
\label{fig:hBN}
\end{figure}

\subsection{Environmental screening}

So far, we have presented results for tBLG suspended in air ($\epsilon_\mathrm{bg}=1$). In experiments, however, the tBLG is placed on or sandwiched by a dielectric substrate (typically, hBN) and the presence of this dielectric environment screens the interaction between electrons in the tBLG~\cite{NAT_I,NAT_S,TSTBLG,SOM,NAT_SS,NAT_MEI,NAT_CO,PHD_3}. In transport experiments, the dielectric substrate separates the tBLG from a metallic gate which is used to control the charge density in the tBLG and the presence of gates further modifies the effective interaction between the electrons in the tBLG [Eq.~\eqref{WMG}].  

Figure~\ref{fig:hBN}(a) compares Hartree band structures of electron-doped tBLG ($\nu=2$) at $\theta=1.54\degree$ (similar band width to that of the experiments in Ref.~\citenum{NAT_MEI}) with $\epsilon_\mathrm{bg}=1$ (tBLG suspended in air) and $\epsilon_\mathrm{bg}=3.9$ (tBLG sandwiched by thick layers of hBN). Surprisingly, the difference between the two band structures is small on the scale of the band width of the flat bands (similar band widths to those in experiments too). To understand this finding, we analyze the Hartree potentials of the two systems. Fig.~\ref{fig:hBN}(b) shows $\Delta V_\mathrm{H}$ (the difference between the Hartree potential in the centers of the AA and AB regions) as a function of doping for the two cases. While one might naively expect that the slope of $\Delta V_\mathrm{H}$ should be reduced by a factor of $\epsilon_\mathrm{bg}=3.9$ when the dielectric environment is included, we find that the reduction is much smaller ($\Delta V_\mathrm{H}$ is only reduced by 30\% when the dielectric environment is included).


The inclusion of metallic gates on both sides of hBN-encapsulated tBLG at a distance of $10$~nm for a twist angle of $1.54\degree$ also has little effect on the band structure [Fig.~\ref{fig:hBN}(a), purple dash-dotted line] because the Hartree potential does not change significantly, as shown in Fig.~\ref{fig:hBN}(b). It is worth noting that most experiments use larger gate distances than 10~nm which would result in an even smaller effect. Very recently, experiments employing very small gate distances reported dramatic changes of the electronic phase diagram and suggested that these were induced by changes in the environmental screening~\cite{IIS}. While further work is required to study the effect of metallic gates for small gate distances and twist angles very close to the magic angle, we stress that the phase diagram is determined by the relative stability of the competing phase, i.e. the total energy differences. It is possible that relatively small changes in the dielectric screening can change the relative stability of the competing phases and thereby lead to drastic changes in the phase diagram, while only mildly affecting quasiparticle properties.



This surprising robustness of the Hartree band structure of tBLG towards changes in the dielectric environment has two reasons. First, the weakening of the Coulomb repulsion by the dielectric substrate allows for a greater inhomogeneity of the charge density. This results in a larger Hartree potential energy than the one that would have been obtained if the charge density had been frozen in its unscreened configuration. Second, the change in the dielectric environment only leads to small changes in the total screening response because the internal screening of the tBLG is already quite strong~\cite{PHD_2,CCRPA}.

\begin{figure*}[t!]
\centering
\begin{subfigure}{0.45\textwidth}
  \includegraphics[width=1\linewidth]{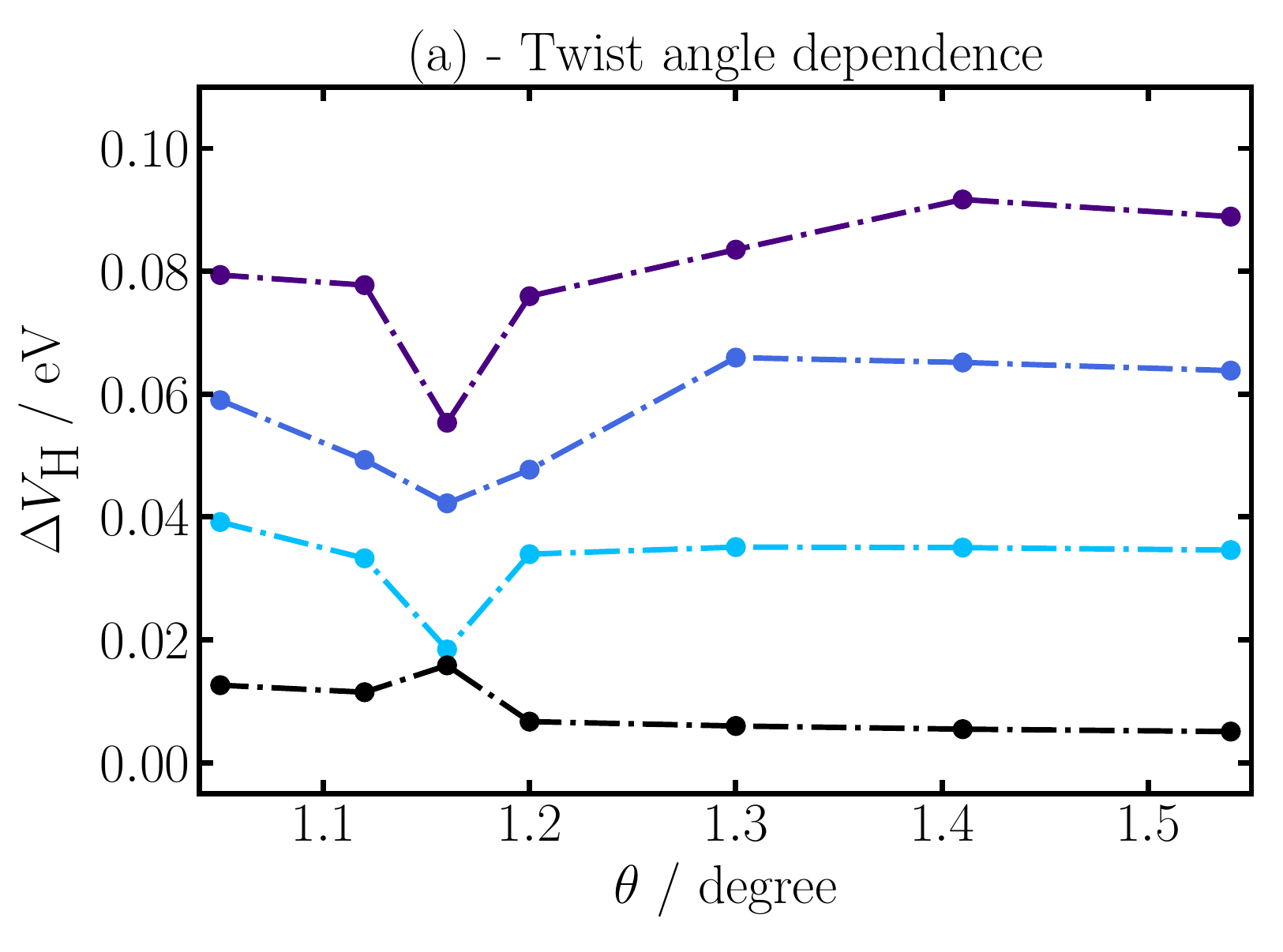}
\end{subfigure}
\begin{subfigure}{0.45\textwidth}
  \includegraphics[width=1\linewidth]{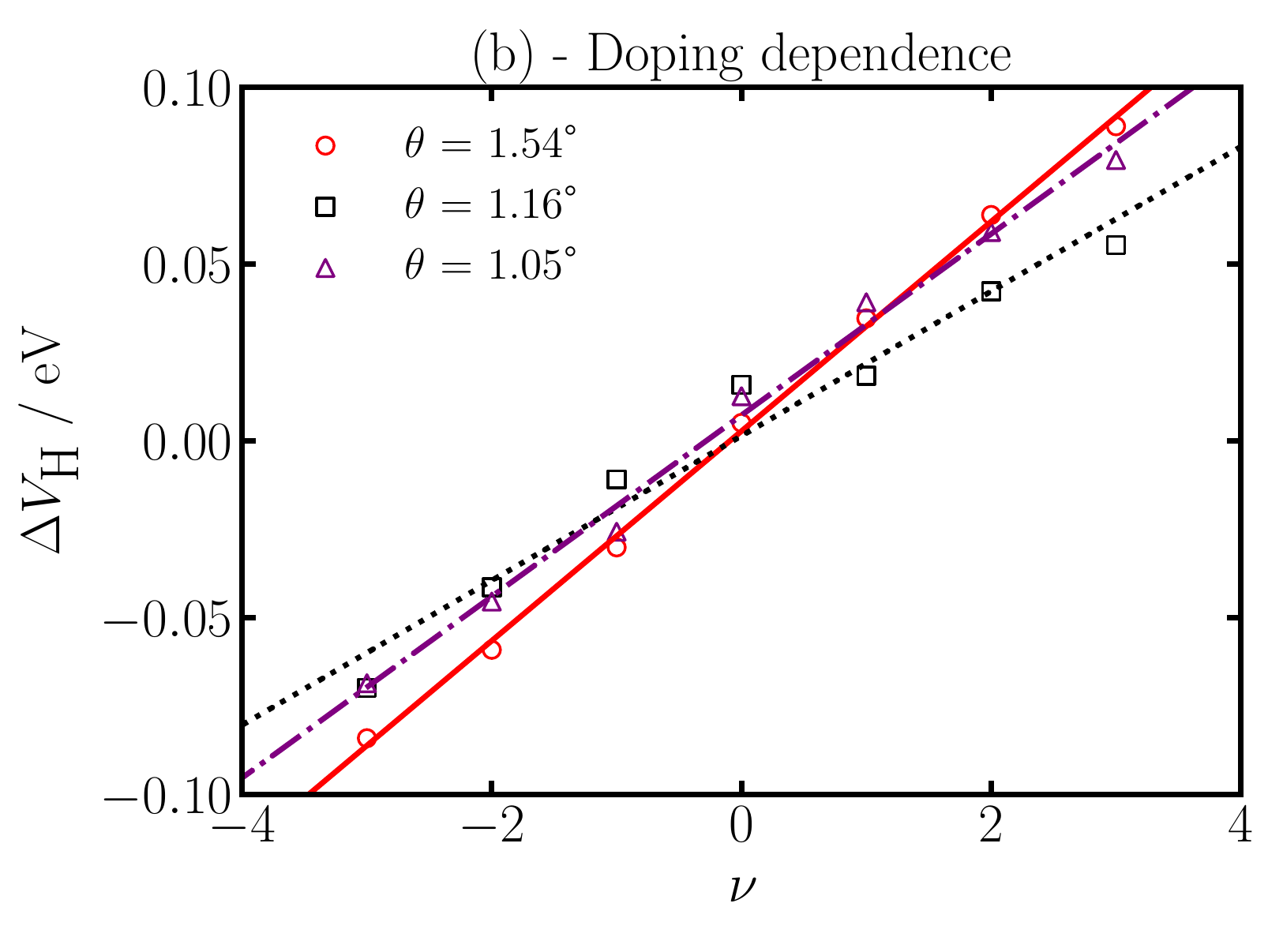}
\end{subfigure}
\caption{(a) Hartree potential difference $\Delta V_\mathrm{H}$ between the AA and AB region as a function of twist angle for undoped (black) and electron-doped twisted bilayer graphene for $\nu=1$ (cyan), $\nu=2$ (blue) and $\nu=3$ (purple). (b) $\Delta V_\mathrm{H}$ as function of doping for three twist angles near the magic angle and linear fits obtained from Eq.~\eqref{FHP}.}
\label{fig:VH}
\end{figure*}

\begin{table}
\centering
\begin{tabular}{ccccc}
\hline
$\theta\degree$   & $\quad$ & V($\theta$) / meV &$\quad$& $\nu_{0}(\theta)$ \\
\hline
\hline
$1.54$      &  & $6.57$           & & $-0.080$  \\
$1.41$      &  & $6.63$           & & $-0.087$  \\
$1.30$      &  & $5.98$           & & $-0.174$  \\
$1.20$      &  & $5.23$           & & $-0.192$  \\
$1.16$      &  &  ---             & &  ---      \\
$1.12$      &  & $5.22$           & & $-0.236$  \\
$1.05$      &  & $5.51$           & & $-0.240$  \\
\hline
\end{tabular}
\caption{Coefficients for the Hartree potential fit, Eq.~\eqref{FHP}. The magic angle cannot be accurately reproduced with this fit so we do not provide parameters here.}
\label{tab:PAR}
\end{table}

In Fig.~\ref{fig:VH} we show how $\Delta V_{\textrm{H}}$ changes as a function of twist angle and doping. It can be seen that there is little change as a function of twist angle (with the exception of the magic angle) and that $\Delta V_{\textrm{H}}$ changes approximately linearly with doping. Except at the magic angle, the doping and twist-angle dependent atomistic Hartree potential energy is accurately described by
\begin{equation}
V_{\textrm{H}}(\textbf{r}) \approx V(\theta) (\nu - \nu_{0}(\theta))\sum_{j=1,2,3}\cos(\textbf{G}_j \cdot \mathbf{r}),
\label{FHP}
\end{equation}

\noindent where $\nu_0(\theta)$ is the doping level where the Hartree potential vanishes, $V(\theta)$ is a twist angle dependent energy parameter and $\textbf{G}_j$ denote the three reciprocal lattice vectors that are used to describe the out-of-plane corrugation of tBLG in Ref.~\citenum{MLWO}. Also note that the form of this equation is very similar to the continuum model of Ref.~\citenum{H_CM}. Table~\ref{tab:PAR} shows the optimal values of these parameters for the twist angles that we have studied and Fig.~\ref{fig:VH}(b) compares the fit to the calculated Hartree potential as function of doping for different twist angles. Using Eq.~\eqref{FHP} as an on-site energy in a tight-binding calculation allows the determination of Hartree-theory band structures without the need for self-consistent calculations. We believe that this approach is a useful starting point for understanding broken symmetry phases in doped tBLG. 


\section{Conclusion}

We have calculated quasiparticle properties, such as band structures and (local) densities of states, of interacting electrons in twisted bilayer graphene as function of doping and twist angle using atomistic Hartree theory. We find that doping results in significant changes to quasiparticle properties which are not captured by tight-binding approaches. In particular, we find that the partially occupied bands flatten between $\Gamma$ and $M$ in the Brillouin zone and even invert upon doping. The resulting local densities of states are in good agreement with several recent scanning tunneling spectroscopy experiments: in particular, we capture the Fermi level pinning and the shapes of the van Hove singularities in the AA regions of tBLG that were reported in these experiments. We predict that the band flattening gives rise to a strong enhancement of the peak in the AB regions. We also study the dependence of quasiparticle properties on the dielectric environment and find that they are surprisingly robust as a consequence of the strong internal screening of tBLG. As a consequence, the properties of broken symmetry phases of tBLG could result from a delicate interplay of long-ranged Coulomb interactions arising from the emergent moir\'e lattice and short-ranged atomic Hubbard interactions inherited from the untwisted bilayer. This will be the subject of future work.  

\section{Acknowledgements}
We wish to thank K. Atalar, P. Guinea, N. Walet, D. Kennes and F. Corsetti for helpful discussions. We also wish to thank A. Kerelsky and L. Xian for sharing their data and for helpful discussions. ZG was supported through a studentship in the Centre for Doctoral Training on Theory and Simulation of Materials at Imperial College London funded by the EPSRC (EP/L015579/1). We acknowledge funding from EPSRC grant EP/S025324/1 and the Thomas Young Centre under grant number TYC-101.

\bibliographystyle{apsrev4-1}
\bibliography{H}

\setcounter{figure}{0}
\setcounter{equation}{0}

\onecolumngrid

\renewcommand{\theequation}{S\arabic{equation}}
\renewcommand{\thefigure}{S\arabic{figure}}

\section{Supplementary Material}

\subsection{Hartree comparison to tight-binding band structures}

\begin{figure*}[h]
\centering
\begin{subfigure}{0.325\textwidth}
  \includegraphics[width=1\linewidth]{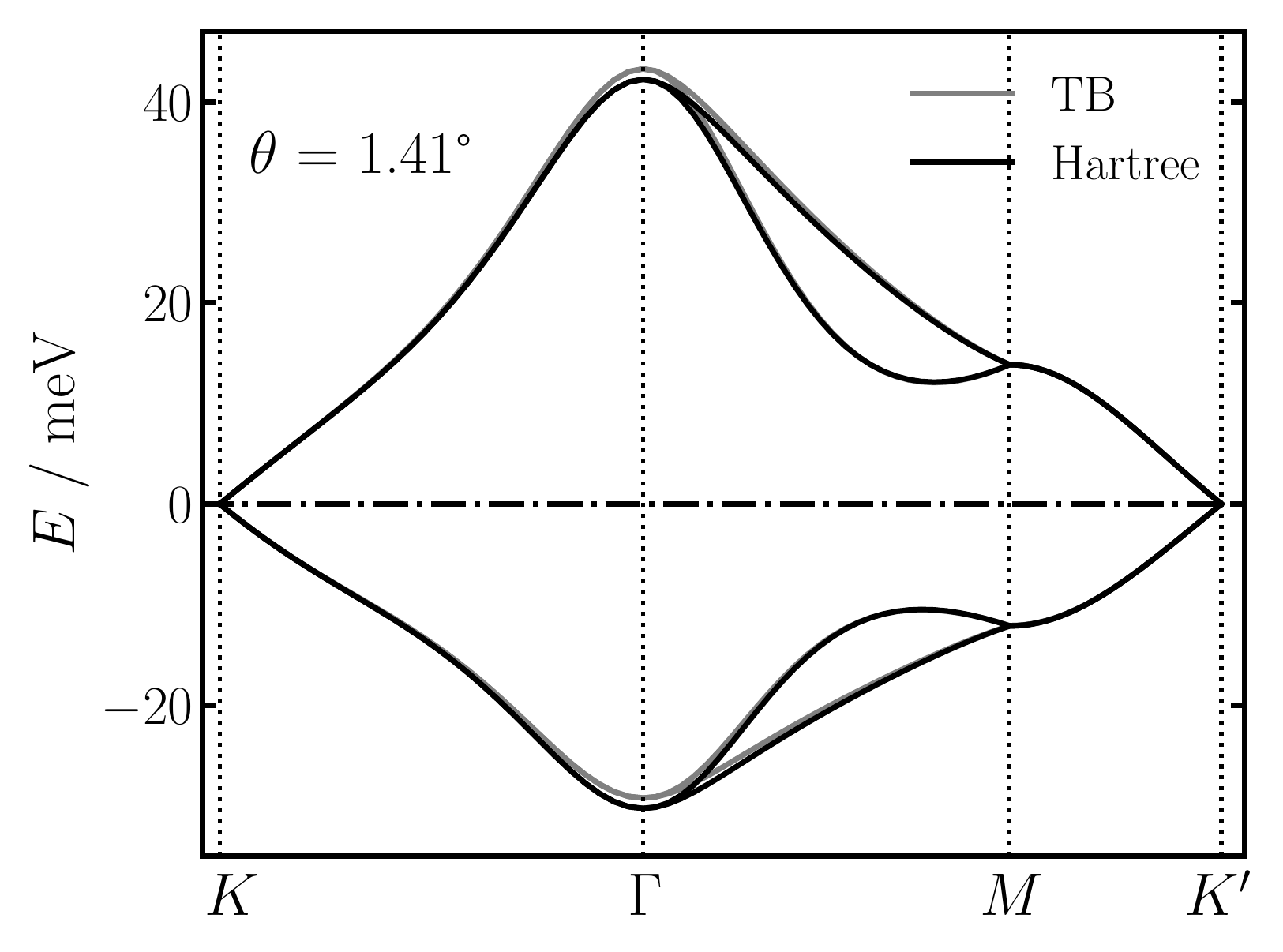}
\end{subfigure}
\begin{subfigure}{0.325\textwidth}
  \includegraphics[width=1\linewidth]{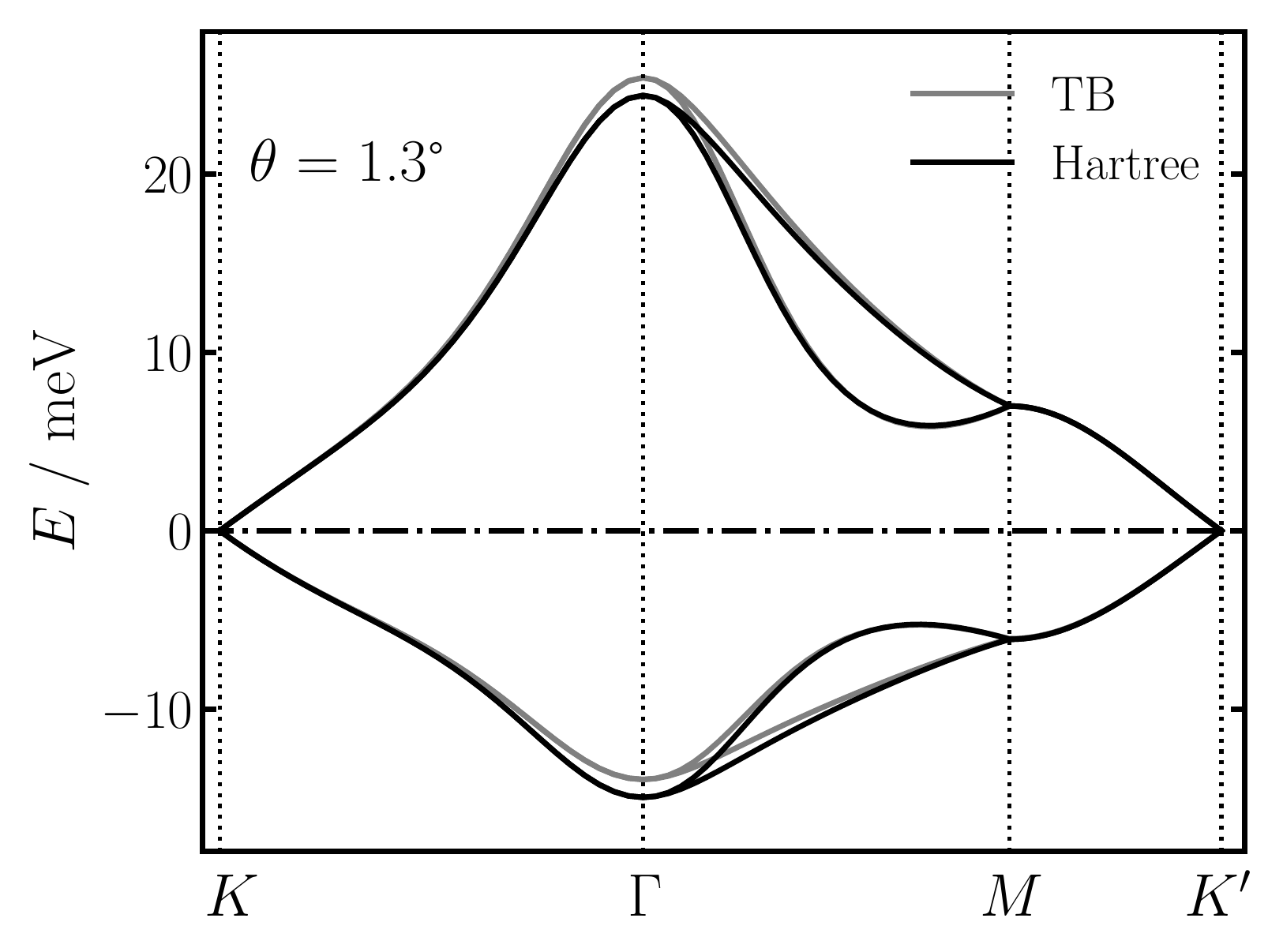}
\end{subfigure}
\begin{subfigure}{0.325\textwidth}
  \includegraphics[width=1\linewidth]{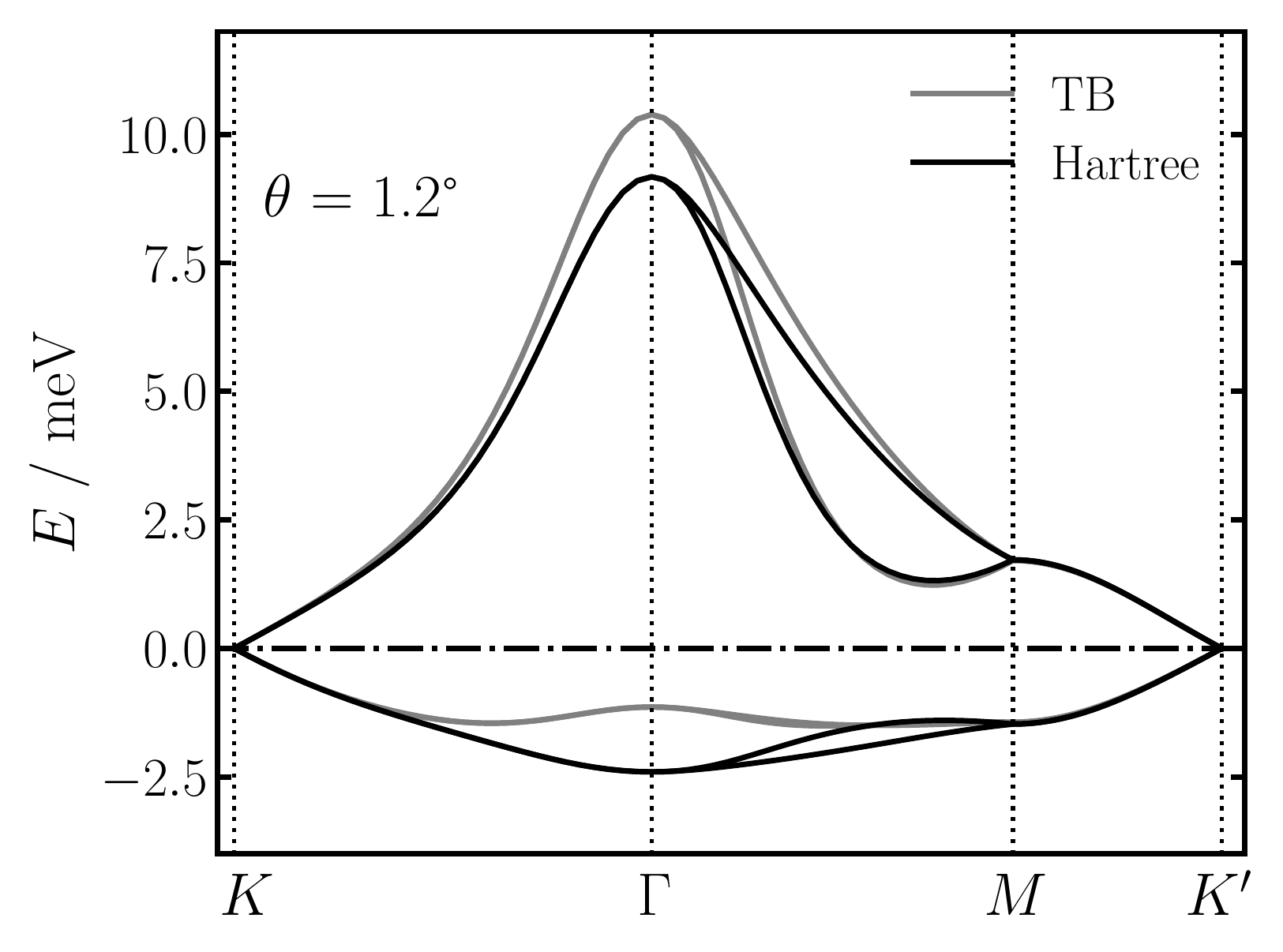}
\end{subfigure}
\begin{subfigure}{0.325\textwidth}
  \includegraphics[width=1\linewidth]{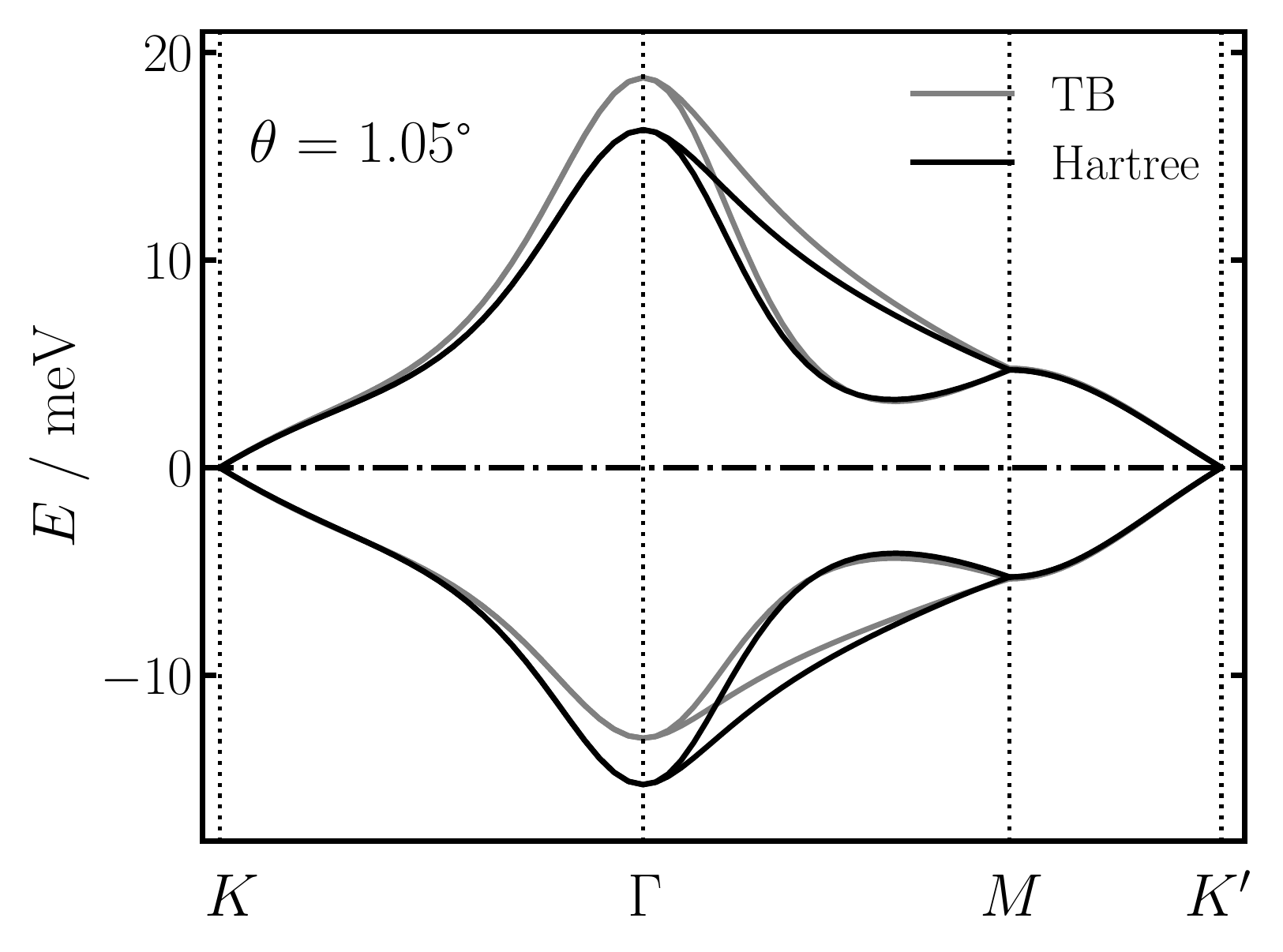}
\end{subfigure}
\begin{subfigure}{0.325\textwidth}
  \includegraphics[width=1\linewidth]{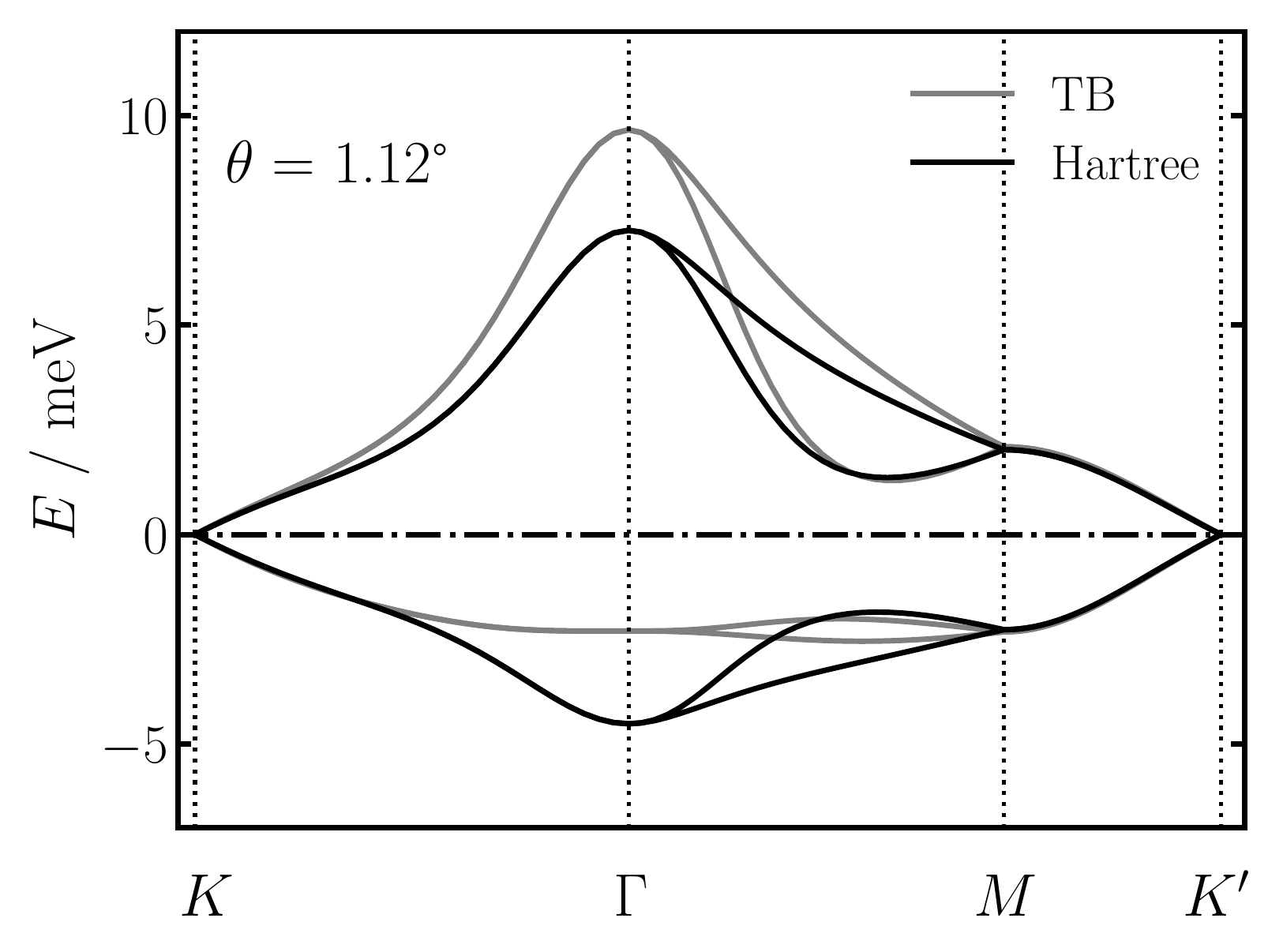}
\end{subfigure}
\begin{subfigure}{0.325\textwidth}
  \includegraphics[width=1\linewidth]{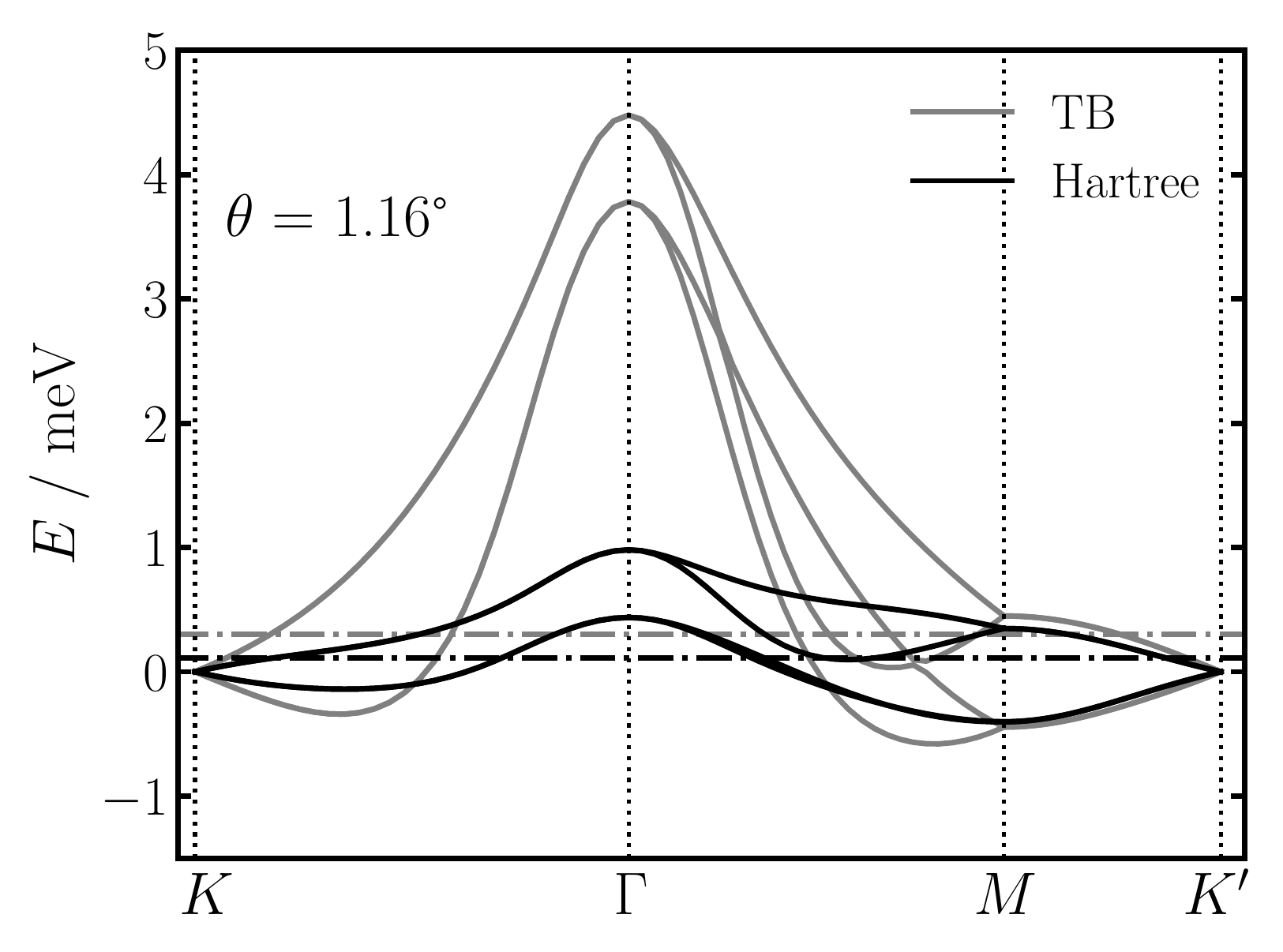}
\end{subfigure}
\caption{Flat band structure along high symmetry path for various twist angles with the Hartree theory and tight-binding. All plots have been aligned at the $K$-point for clarity such that the energy, $E$, is relative to the $K$-point, and the energy scale of each plot is different to clearly show the flat band structures. Dotted-dashed horizontal lines denote the Fermi energy. The band distortions from the Hartree interaction are of 5-10 meV at charge neutrality, which can be significant right at the magic angle.}
\label{Sfig:BS_NP}
\end{figure*}


\subsection{Additional band structures}

\begin{figure*}[h]
\centering
\begin{subfigure}{0.35\textwidth}
  \includegraphics[width=1\linewidth]{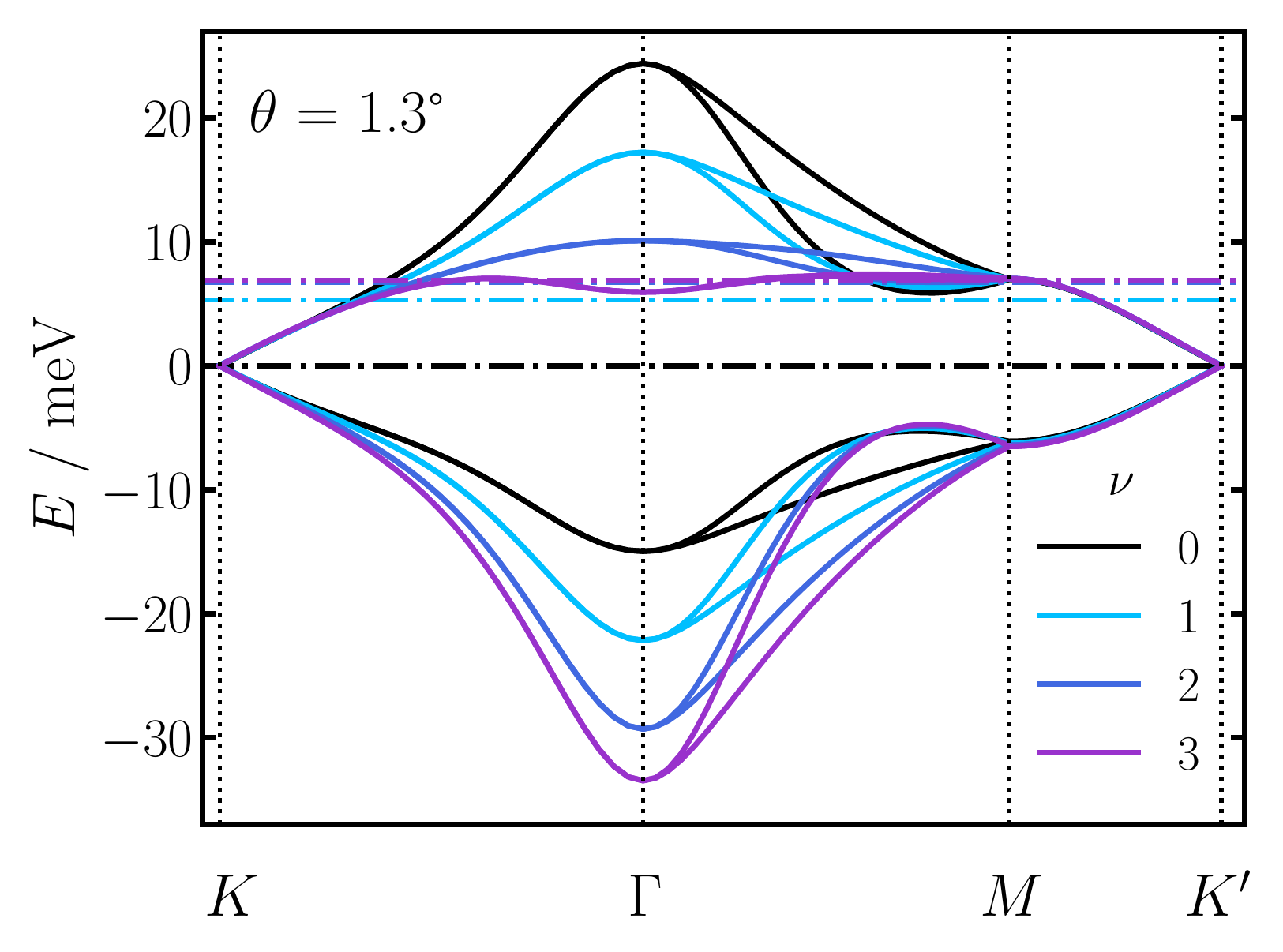}
\end{subfigure}
\begin{subfigure}{0.35\textwidth}
  \includegraphics[width=1\linewidth]{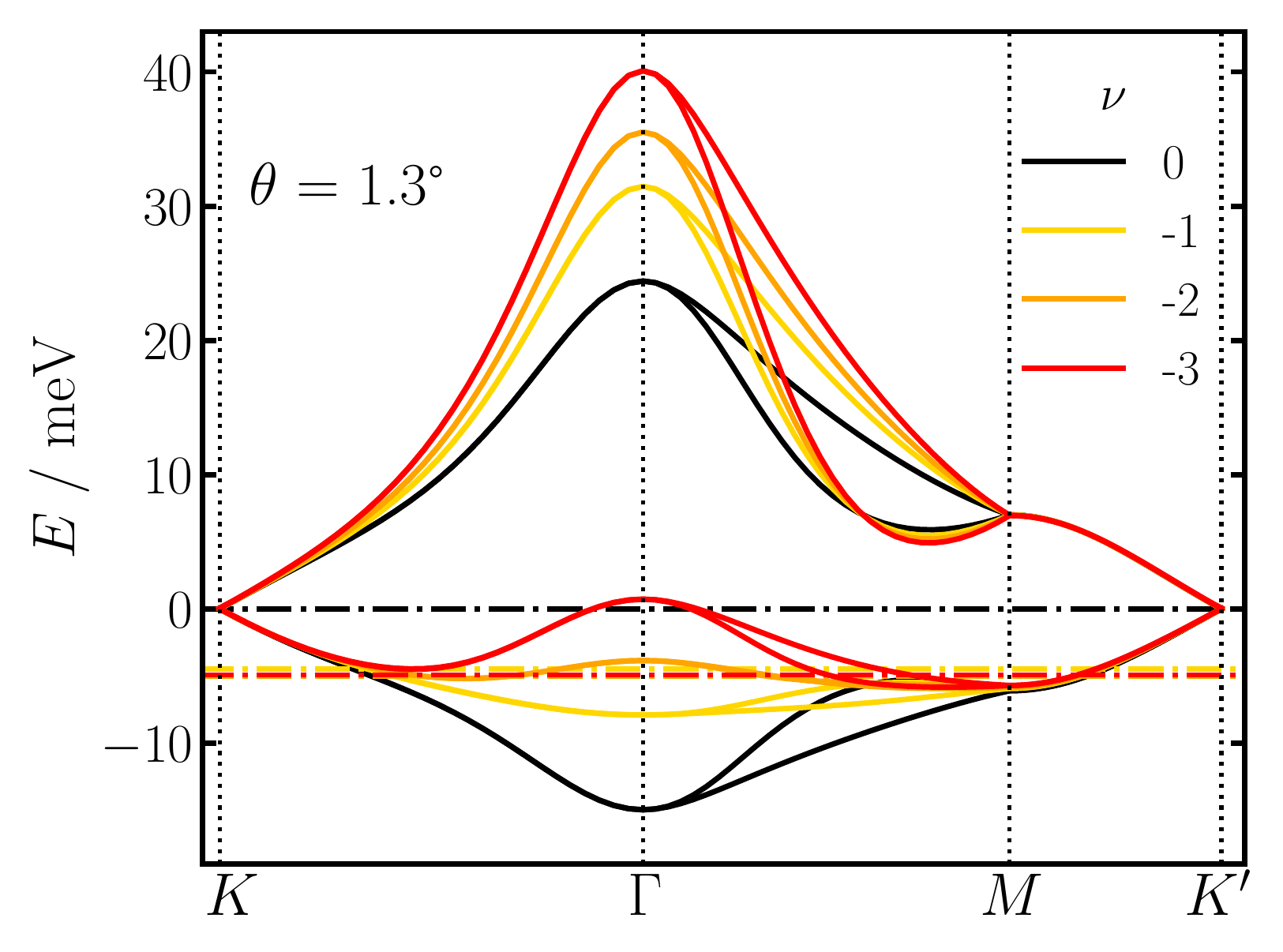}
\end{subfigure}
\caption{Flat band structure along high symmetry path for a twist angle of $\theta = 1.3\degree$ with electron (left) and hole doping (right). Same format as Fig.~\ref{Sfig:BS_NP}.}
\label{Sfig:BS_FIN}
\end{figure*}

\begin{figure*}[h]
\centering
\begin{subfigure}{0.45\textwidth}
  \includegraphics[width=1\linewidth]{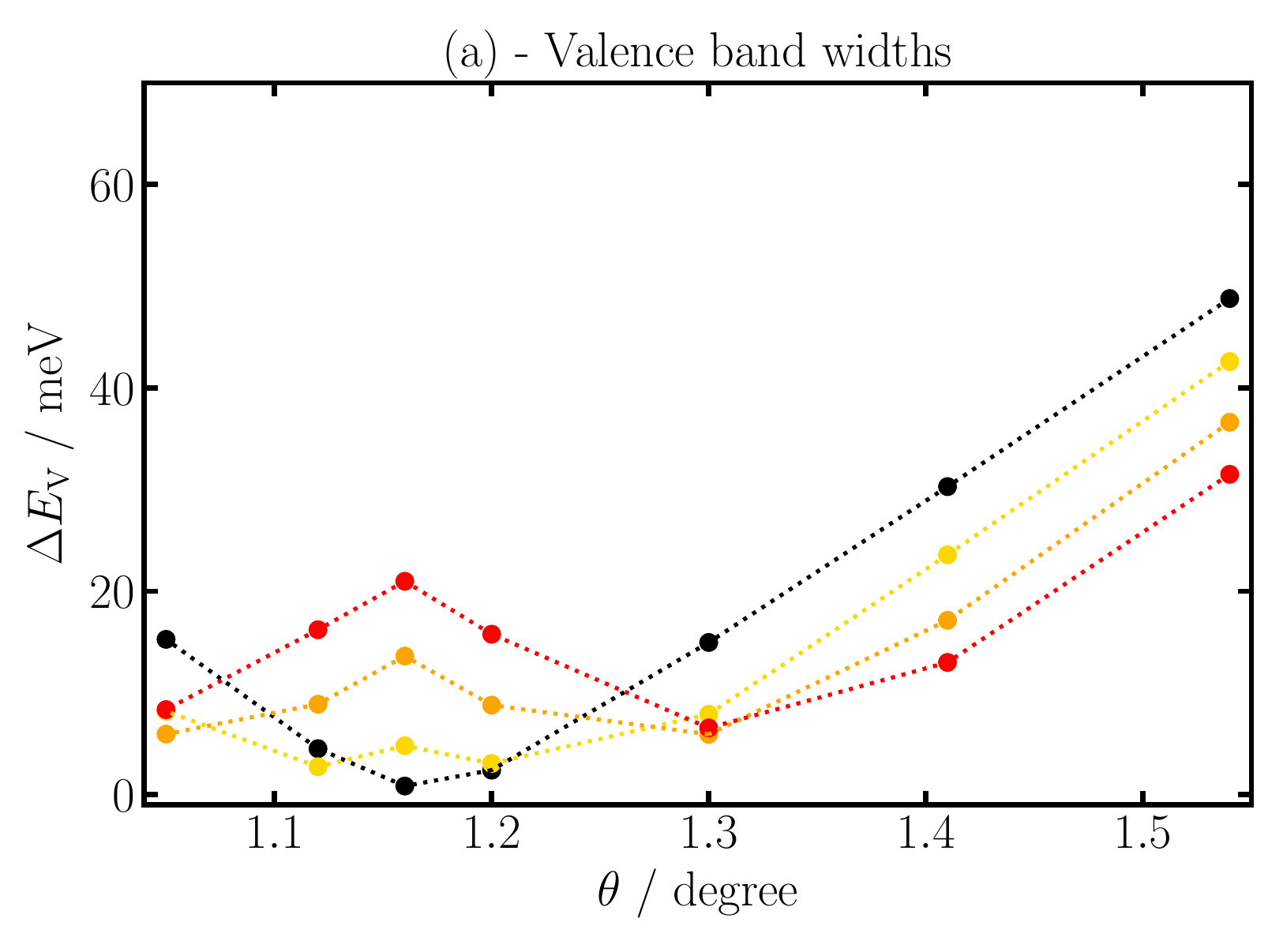}
\end{subfigure}
\begin{subfigure}{0.45\textwidth}
  \includegraphics[width=1\linewidth]{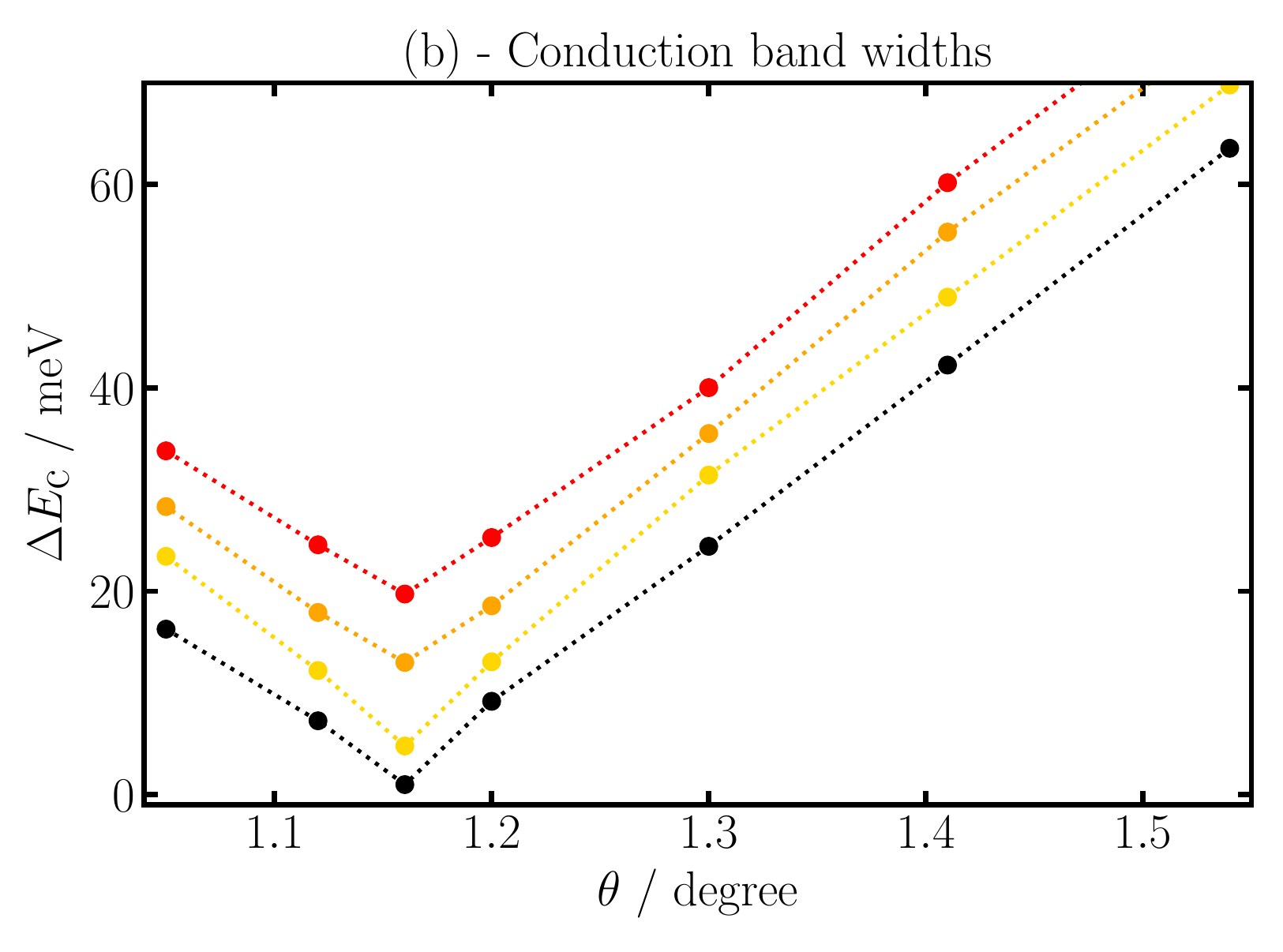}
\end{subfigure}
\caption{Valence (a) and conduction (b) band widths as a function of twist angle for charge neutrality and hole doped systems.}
\label{Sfig:BWh}
\end{figure*}
\clearpage
\newpage

\subsection{Additional DOS calculations}

\begin{figure*}[h]
\centering
\begin{subfigure}{0.33\textwidth}
  \includegraphics[width=1\linewidth]{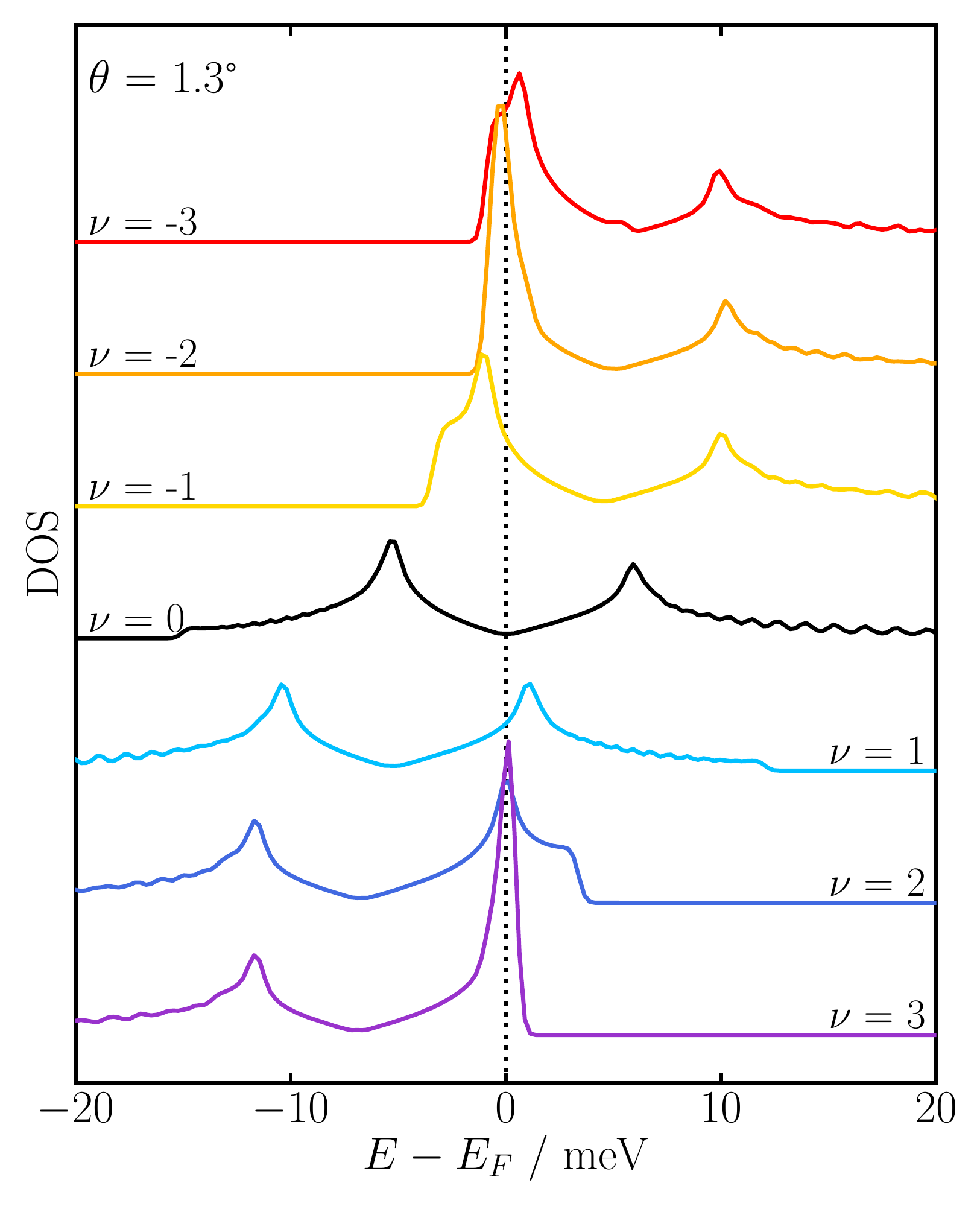}
\end{subfigure}
\begin{subfigure}{0.33\textwidth}
  \includegraphics[width=1\linewidth]{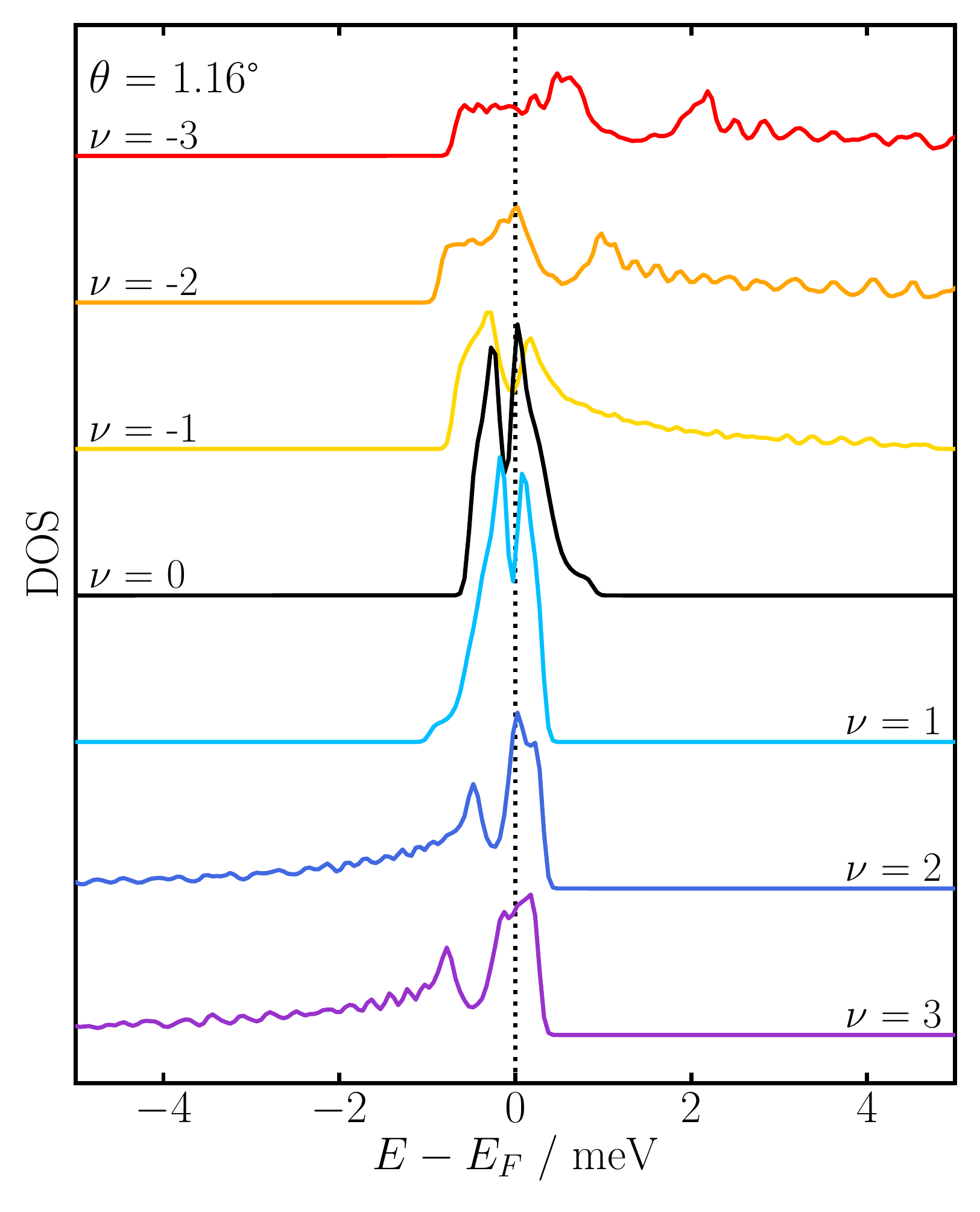}
\end{subfigure}
\begin{subfigure}{0.33\textwidth}
  \includegraphics[width=1\linewidth]{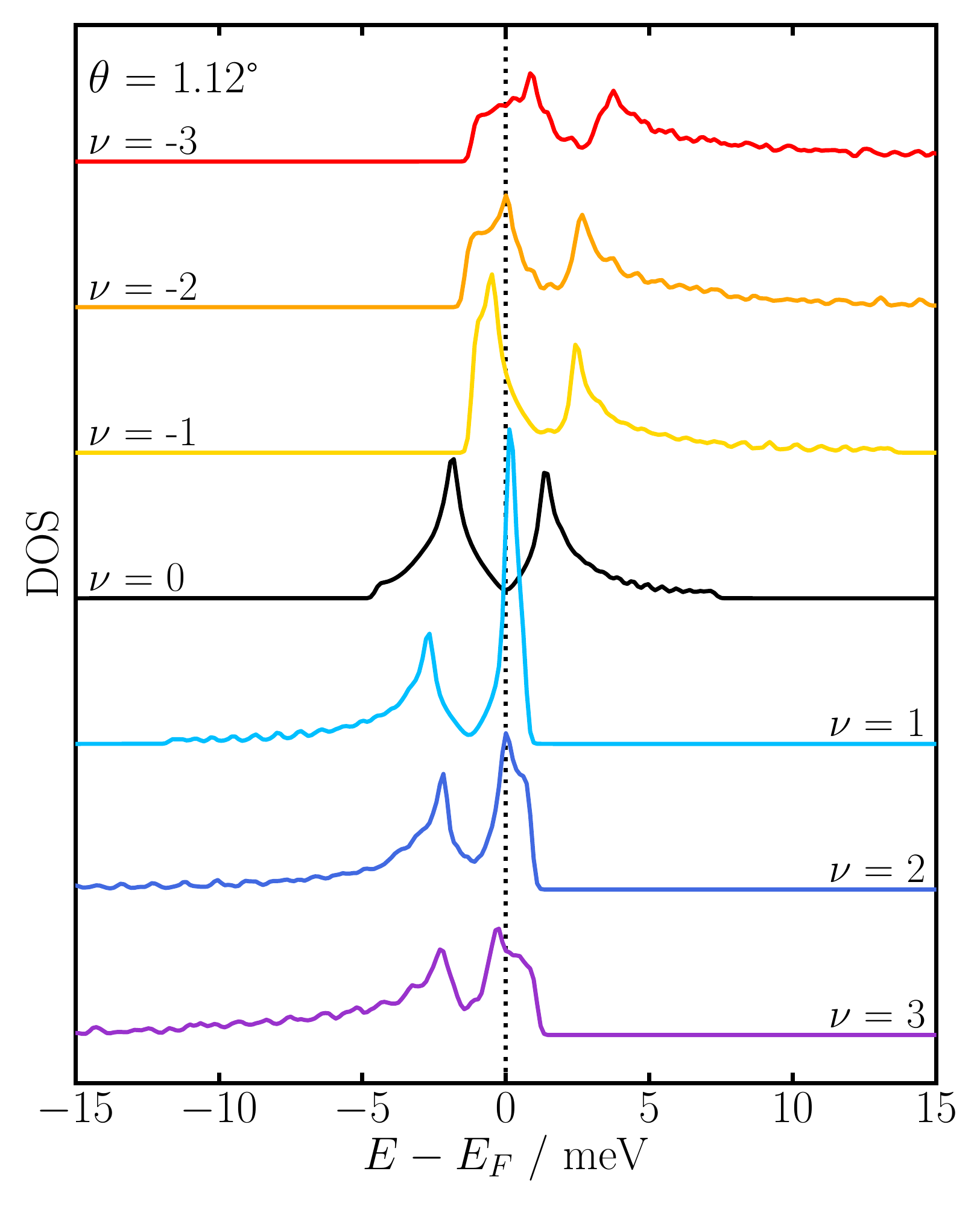}
\end{subfigure}
\begin{subfigure}{0.33\textwidth}
  \includegraphics[width=1\linewidth]{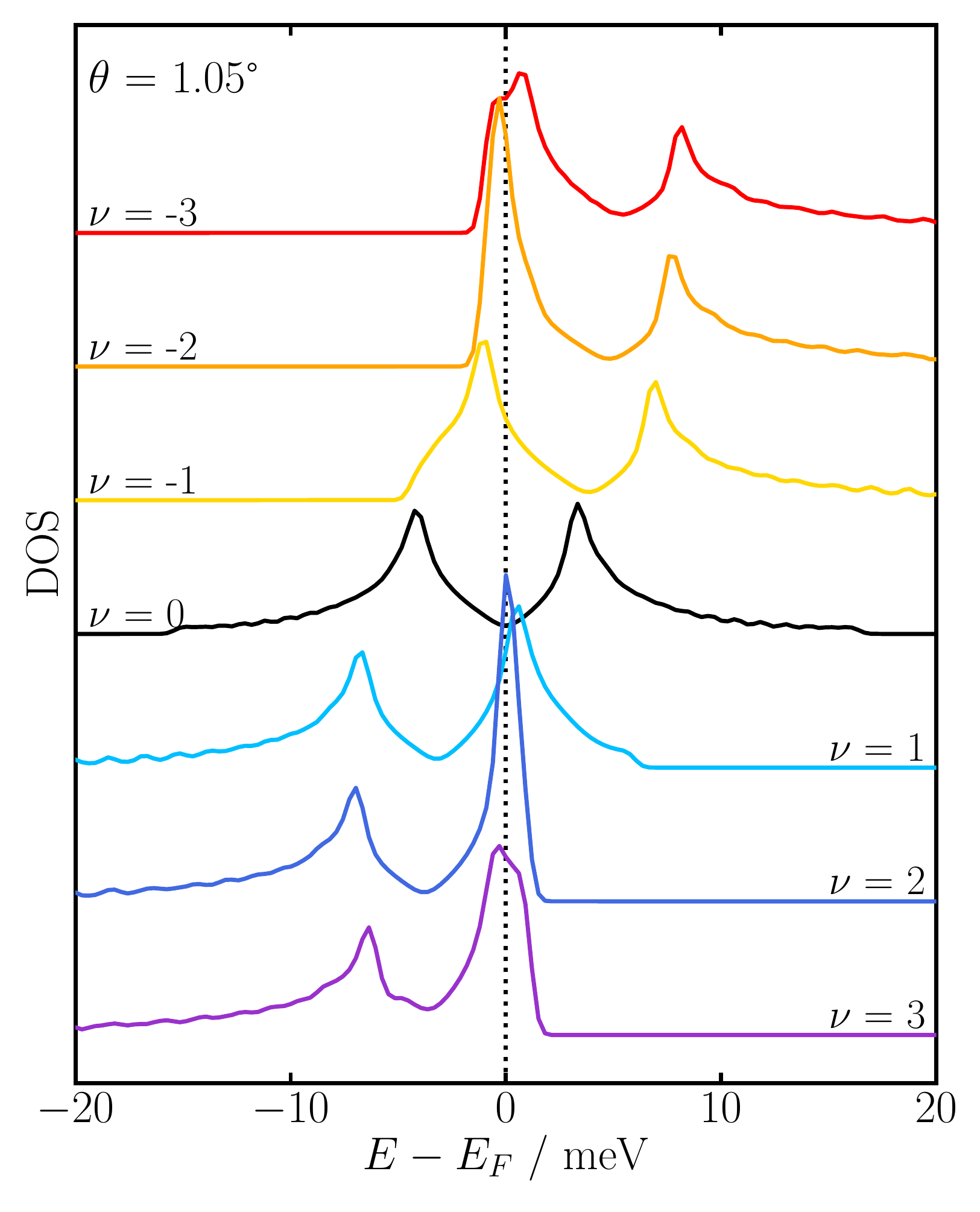}
\end{subfigure}
\caption{Density of states as a function of energy relative to the Fermi energy, for various twist angle and doping levels.}
\label{Sfig:DOS}
\end{figure*}

\newpage

\subsection{Additional LDOS calculations}

\begin{figure*}[h]
\centering
\begin{subfigure}{0.33\textwidth}
  \includegraphics[width=1\linewidth]{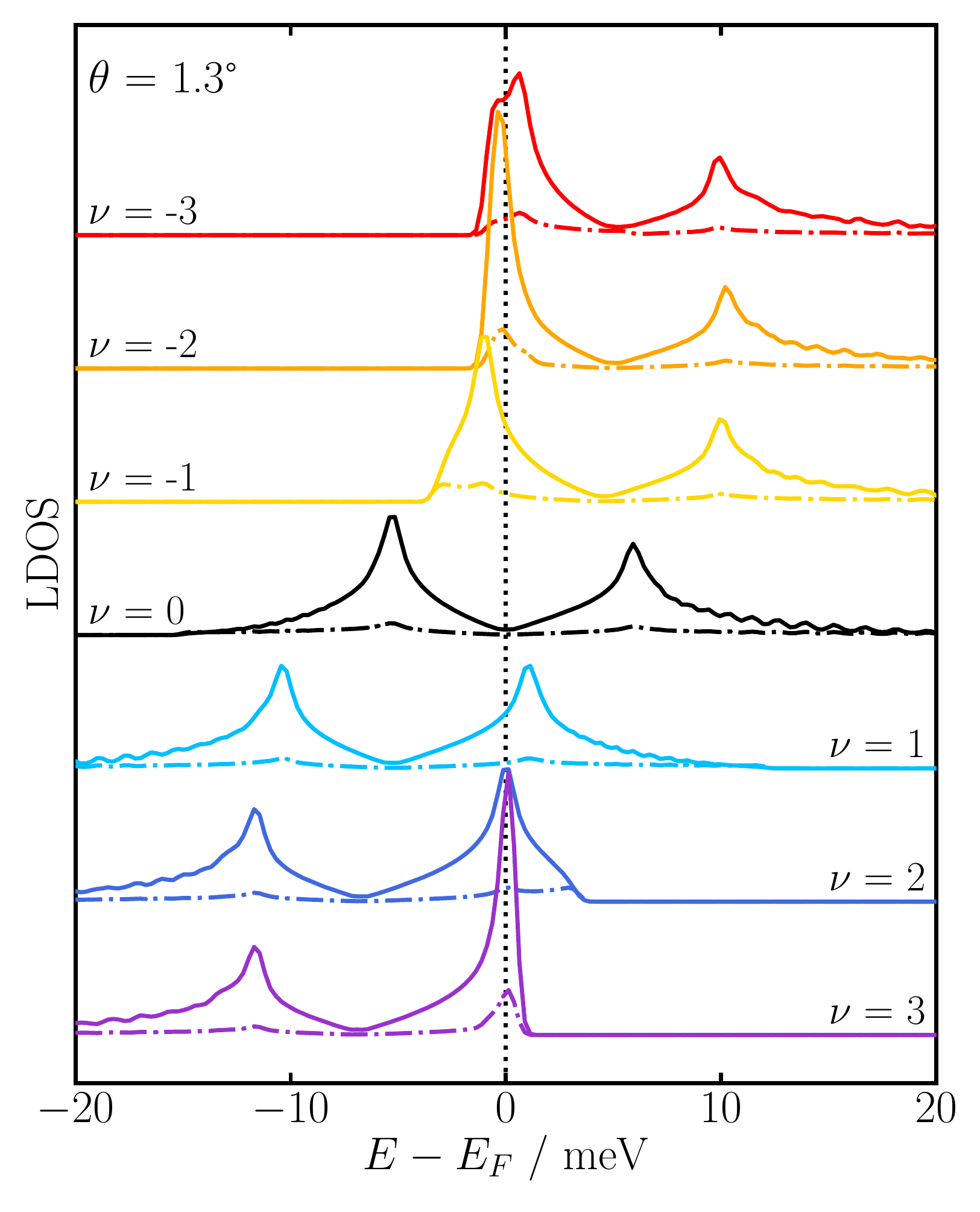}
\end{subfigure}
\begin{subfigure}{0.33\textwidth}
  \includegraphics[width=1\linewidth]{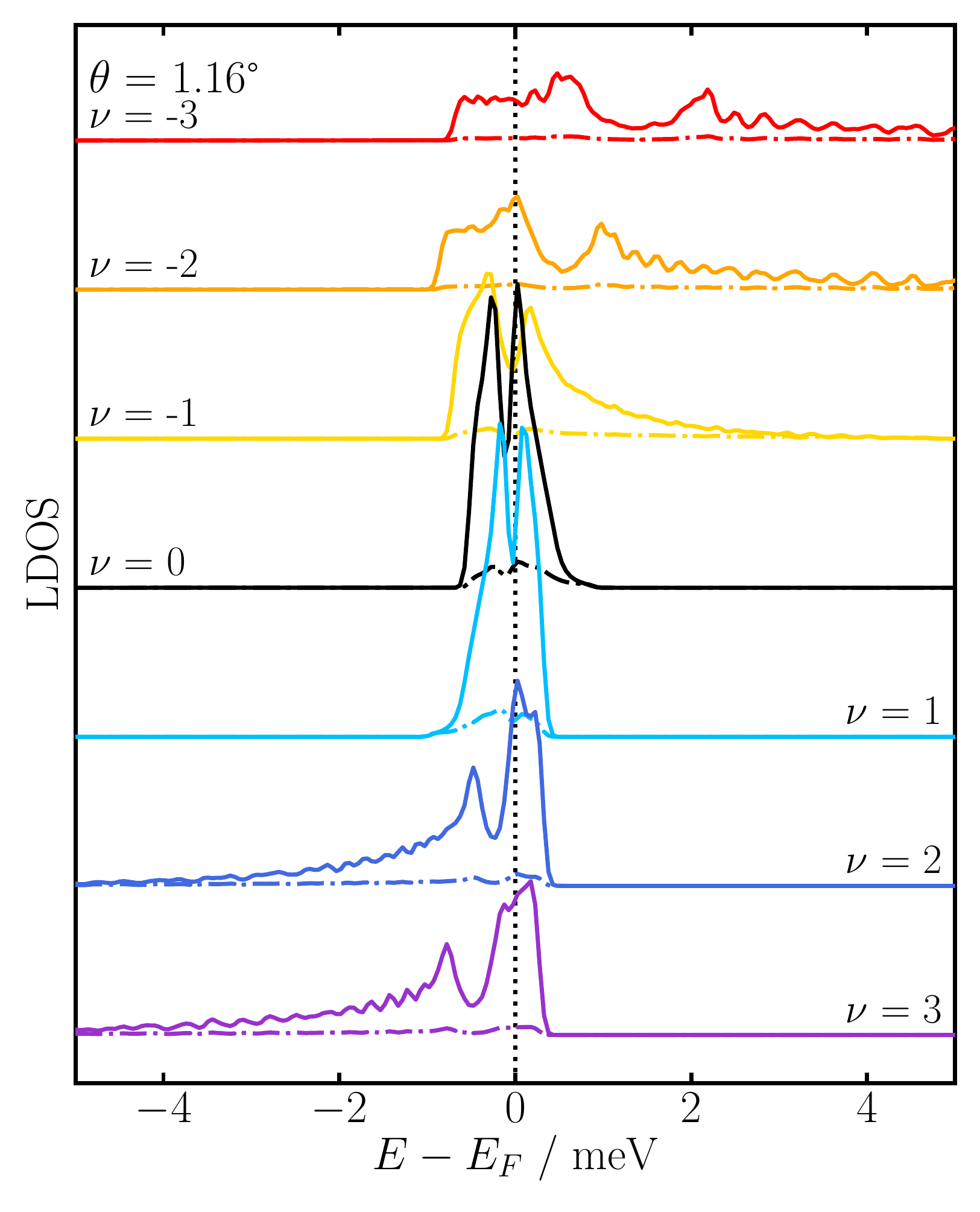}
\end{subfigure}
\begin{subfigure}{0.33\textwidth}
  \includegraphics[width=1\linewidth]{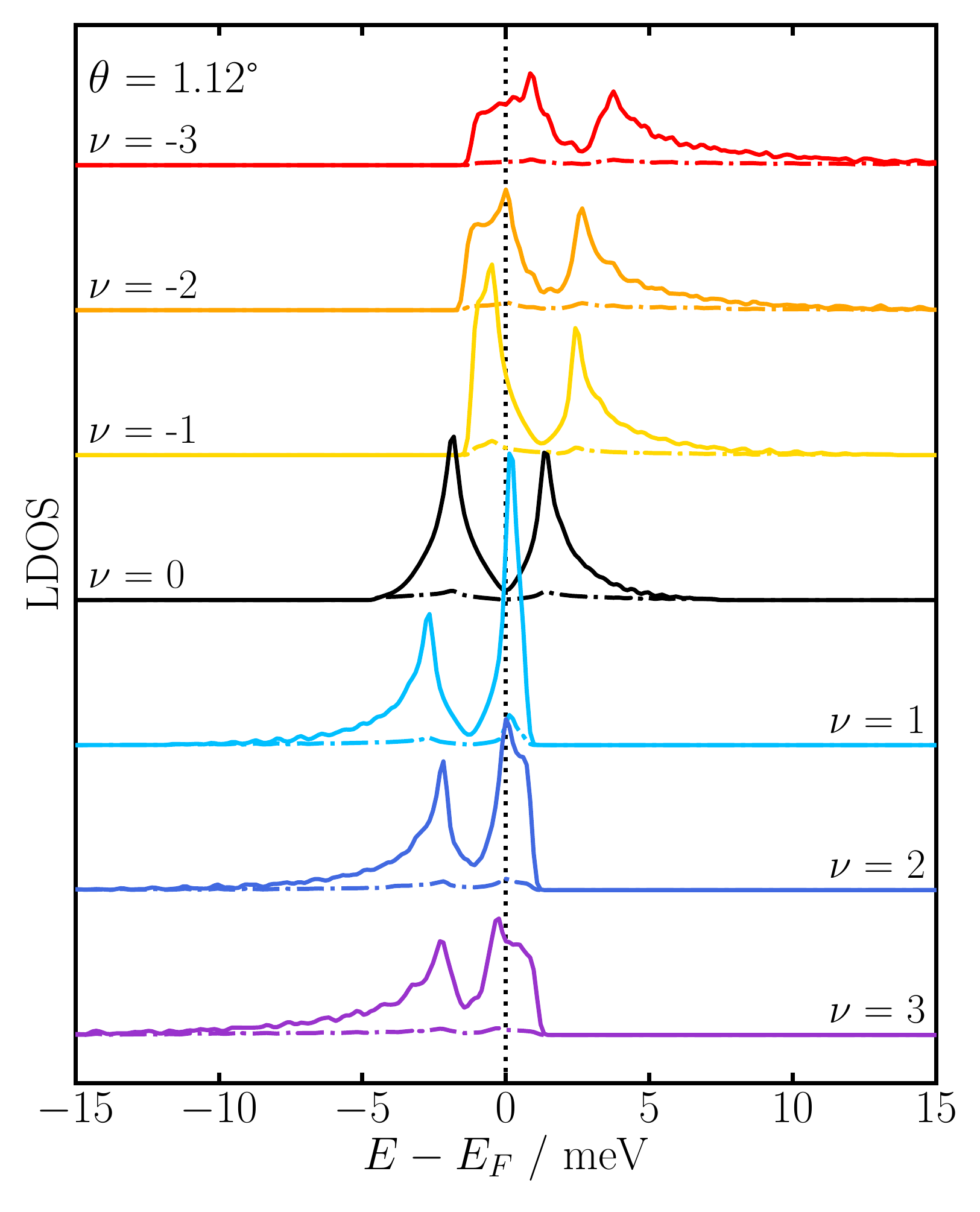}
\end{subfigure}
\begin{subfigure}{0.33\textwidth}
  \includegraphics[width=1\linewidth]{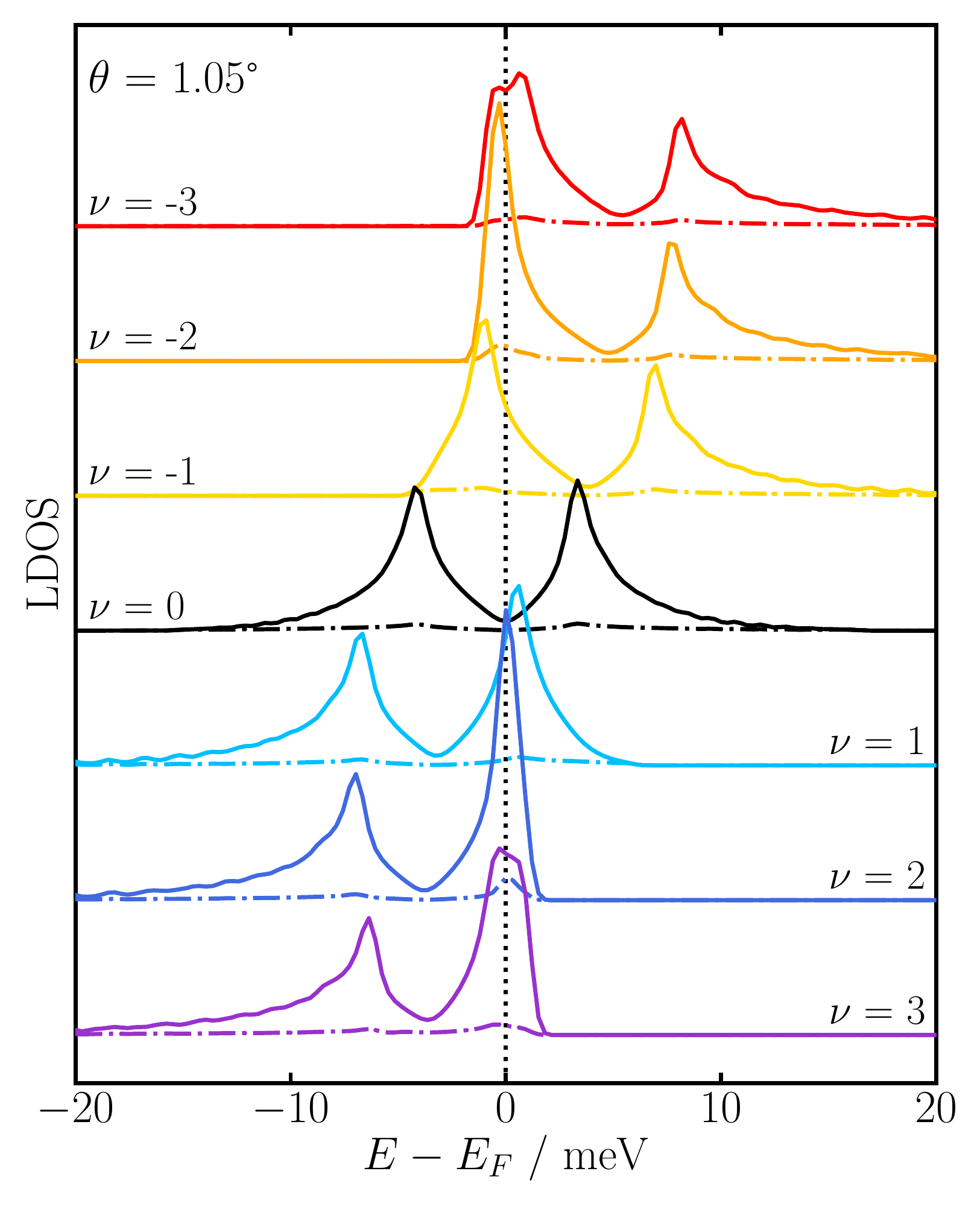}
\end{subfigure}
\caption{Local density of states in the AA (solid curve) and AB (dotted-dashed) region for various twist angle and doping levels. }
\label{Sfig:LDOS}
\end{figure*}

\end{document}